%% file: GM_2013_v_03.tex
\documentclass{PoS}

\usepackage{amsmath,amssymb}

\def\e{{\,\rm e}} 
\def\Li{{\,\rm Li}}
\def\d{{\,\rm d}}
\def\I{{\,\rm Im}}

\title{A new determination of the charm mass from the non-analytic reconstruction of the heavy quark correlator}

\ShortTitle{Charm mass and non-analytic reconstruction}

\author{\speaker{David Greynat}\thanks{This work has been supported by the Spanish DGIID-DGA grant 2009-E24/2, the Spanish  MICINN grants FPA2009-09638 and CPAN-CSD2007-00042 and by the Deutsche Forschungsgemeinschaft DFG through the Collaborative Research Center "The Low-Energy Frontier of the Standard Model" (SFB 1044).}\\
Departamento de F\'{i}sica Te\'orica, Facultad de Ciencias\\
Universidad de Zaragoza\\
50009 Zaragoza, Spain\\
E-mail: \email{david.greynat@gmail.com}}

\author{Pere Masjuan\\
Institut f\"{u}r Kernphysik, \\
Johannes Gutenberg-Universit\"{a}t\\
J.J. Becher-Weg 45\\
D-55099 Mainz, Germany\\
E-mail: \email{masjuan@kph.uni-mainz.de}}

\abstract{Using the new non-analytic reconstruction method obtained from Mellin-Barnes properties, one can extract the value $m_c(\overline{\text{MS}}) = 1.12 \pm 0.08 \;\; \text{GeV}$ from experimental data of the radiation-corrected measured hadronic cross section to the calculated lowest-order cross section for muon pair production in the heavy-quark approximation.}

\FullConference{Xth Quark Confinement and the Hadron Spectrum,\\
		October 8-12, 2012\\
		TUM Campus Garching, Munich, Germany}

\begin{document}
\section{Introduction} 

An accurate determination of the charm mass plays an important role on the precise physical evaluation of several observables, from K and B decays to CKM matrix elements and in lattice QCD. One of the usual techniques to extract the charm mass is to use the sum rules approach based on the relation between the moments of the production rate $R$ and the  inverse power of the square mass of the $c$ quark, and the Pad\'e method (see \cite{Dehnadi:2011gc,Masjuan:2009wy}). This approach should confront the fact that one have to employ the moments of the integral of R over the whole energy range, which are \textit{global} properties, even though they are only known up to a certain scale $\Lambda$ (since we only know experimentally $R$ in a finite window). We propose to wield the \textit{local} properties of $R$ through a new "non-analytic reconstruction" method \cite{Greynat:2010kx, Greynat:2011zp}. As we will show, this approach allows us to obtain local properties of the heavy quark correlators at each points of the spectrum with a systematic error and then to find a value of the charm mass directly on a $\chi^2$ regression on the experimental points. 

\section{Details of the method} 
\subsection{Non-analytic reconstruction}

Let us consider the vector polarization function 
\begin{equation}
\left(q_{\mu} q_{\nu}- q^2 g_{\mu\nu}\right)\  \Pi(q^2) =  i  \int\!\d^4\,x \e^{iqx}\ \left< 0 \left| \mathrm{T}\,  j_\mu(x) \, j^\mu(0) \right| 0 \right> \;,
\end{equation}
with the current $j_\mu(x)=\overline{\psi}(x) \gamma_\mu \psi (x)$, which has a cut in the complex plane starting at $q^2=4 m^2$, where $m$ is the (pole) mass of the heavy quark considered. In QCD perturbation theory, it can be expanded as 
\begin{equation}
\label{Pi}
\Pi(q^{2}) =\Pi(0) + \Pi^{(0)}(q^{2}) + \left(\frac{\alpha_{s}}{\pi}\right)\Pi^{(1)}(q^{2})+\left(\frac{\alpha_{s}}{\pi}\right)^{2}\Pi^{(2)}(q^{2})\ + \left(\frac{\alpha_{s}}{\pi}\right)^{3}\Pi^{(3)}(q^{2}) + {\mathcal O}(\alpha_s^4 )\;,
\end{equation}
where only $\Pi^{(0)}$ and $ \Pi^{(1)}$ are know analytically, (for $z= q^2/4m^2$)
\begin{equation}
\Pi^{(0)}(z)=\frac{3}{16 \pi^2}\left[\frac{20}{9}+\frac{4}{3\ z}-\frac{4(1-z)(1+2\ z)}{3\ z}\ G(z)\right]\;,
\end{equation}
and
\begin{multline}
\Pi^{(1)}(z) = \frac{3}{16\pi^{2}}\left[\frac{5}{6}+\frac{13}{6z}-\frac{(1-z)(3+2z)}{z}G(z)+
\frac{(1-z)(1-16z)}{6z}G^{2}(z)\right.\\
-\,\left.\frac{(1+2z)}{6z}\left(1+2z(1-z)\frac{d}{dz}\right)\frac{I(z)}{z}\right],
\end{multline} 
in which we used the auxiliary functions, 
\begin{align}
G(z)&=\frac{2\ u\  \log u}{u^2-1}\\
I(z) & = 6\Big[\zeta_{3}+4\,\mbox{Li}_{3}(-u)+2\,\mbox{Li}_{3}(u)\Big] \nonumber\\
&\hspace*{0.8cm} -8\Big[2\,\mbox{Li}_{2}(-u)+\mbox{Li}_{2}(u)\Big]\ln u -2\Big[2\,\ln(1+u)+\ln(1-u)\Big]\ln^{2}u\,,
\end{align}
and
\begin{equation}
u=\frac{\sqrt{1-1/z}-1}{\sqrt{1-1/z}+1}\;.
\end{equation}

As it has been shown \cite{Greynat:2010kx, Greynat:2011zp} even if the functions $\Pi^{(2)}$ and $\Pi^{(3)}$ are unknown analytically, one can reconstruct them from their expansions around $q^2 \rightarrow 0$ (Taylor expansion), $q^2\rightarrow 4m^2$ (threshold expansion) and $q^ 2 \rightarrow \infty$ (OPE), as
\begin{equation}\label{approx}
\Pi^ {(k)}(z) = \sum_{n=0}^{N_k^*} \Omega^{(k)}(n) \omega^n + \sum_{p,\ell} (-)^\ell \left[\alpha_{p,\ell}^{(k)}\Li^ {(\ell)}(p,\omega) -\beta_{p,\ell}^{(k)}\Li^ {(\ell)}(p,-\omega)\right] + \mathcal{E}^{(k)}(N_k^ *,\omega)\;.
\end{equation}   

\begin{figure}[h]
\begin{center}
\includegraphics[width=12cm]{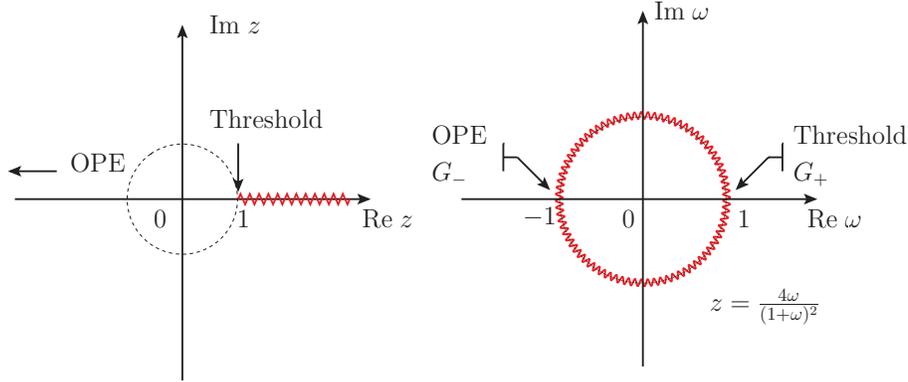} \label{fig:omegaplan}
\end{center}
\caption{Conformal mapping between $z$ and $\omega$.}
\end{figure}

Let emphasize a little this expression. First one defines the so-called \textit{conformal change of variable} 
\begin{equation}\label{conformal}
    z=\frac{4 \omega}{(1+\omega)^2}\qquad , \qquad \omega=\frac{1-\sqrt{1-z}}{1+\sqrt{1-z}}\  .
\end{equation}
This change of variables maps the cut $z$ plane into a unit disc in the $\omega$ plane, as we can see on Figure \ref{fig:omegaplan}. The physical cut $z\in [1, \infty[$ is transformed into the circle $|\omega| = 1$ . The points $z = 0$ into $\omega = 0$, $z =1$ into $\omega=1$, the limit   $z \rightarrow +\infty \pm i \varepsilon$ into $\omega \rightarrow -1 \pm i \varepsilon$, and $z \rightarrow -\infty $ into $\omega \rightarrow -1$.

For both functions  $\Pi^{(2)}$ and $\Pi^{(3)}$, Feynman diagrams calculations at $q^2 \rightarrow 0$ give the expansions  up to an order $N_k^*$ (for $k=2,3$)   
\begin{equation}
\Pi^{(k)}(z)\underset{q^2 \rightarrow 0}{=}\sum_{n=0}^{N_k^*}C^{(k)}(n)z^n + \mathcal{O}\left(z^ {N_k^*+1}\right)\underset{\omega \rightarrow 0}{=} \sum_{n=0}^{N_k^*} \Omega^{(k)}(n) \omega^n+ \mathcal{O}\left(\omega^ {N_k^*+1}\right)\, ,
\end{equation}
where the relation between the two coefficients $C^{(k)}$ and $\Omega^{(k)}(n)$ is
\begin{align}
\Omega^{(k)}(n) &= (-1)^n \sum_{p=1}^{n}  \frac{(-1)^p\;4^p\;\Gamma(n+p)}{\Gamma(2p)\Gamma(n+1-p)}\; C^{(k)}(p) \;,\label{Omega0=C}\\
 C^{(k)}(n)  &=2^{1-2n}\Gamma\left(2n\right)\;\sum_{p=1}^n \frac{ p}{\Gamma\left(1+n-p\right) \Gamma\left(1+n+p\right)} \;  \Omega^{(k)}(p)  \; .\label{C=Omega0}
\end{align}

The main part of the approximation in (\ref{approx}) lies on the combination of the polylogarithms functions,
\begin{equation}
\Li^ {(\ell)}(s,\omega) =\frac{\d^\ell}{\d s^ \ell} \left[\frac{\omega}{\Gamma(s)} \int_0^1 \frac{\d t }{1- \omega t} \log^{s-1}\left(\frac{1}{t}\right) \right]\underset{ |\omega| < 1}{=} (-1)^\ell \sum_{n=1}^\infty \frac{\log^\ell n}{n^s}\;  \omega^n\;, 
\end{equation}  
and the analytic evaluation of the coefficients $\alpha_{p,\ell}^{(k)}$ and $\beta_{p,\ell}^{(k)}$. In order to reconstruct $\Pi^{(2)}$ and $\Pi^{(3)}$, we collect here their corresponding coefficients  (see \cite{Greynat:2010kx, Greynat:2011zp} for more details)
\begin{equation}
\left\{ 
\begin{aligned}
\alpha^{(2)}_{0,0} &= 3.44514 \\
\alpha^{(2)}_{1,0} &=-0.492936 \\
\alpha^{(2)}_{1,1} &= 2.25\\
\alpha^{(2)}_{2, 0} &= 3.05433\\
\end{aligned}\;,
\right.
\hspace{2cm}
\left\{ 
\begin{aligned}
\beta^{(2)}_{1, 0} &= 0.33723\\
\beta^{(2)}_{1, 1} &= 0.211083\\
\beta^{(2)}_{3, 0} &= 0.183422\\
\beta^{(2)}_{3,1}  &= -0.620598\\
\end{aligned}
\right.\;,
\end{equation}

\begin{equation}
\left\{ 
\begin{aligned}
\alpha^{(3)}_{-1,0}&= 10.5456 \\
\alpha^{(3)}_{0,1} &= 31.0063\\
\alpha^{(3)}_{0,0} &= -11.0769 \\
\alpha^{(3)}_{1,0} &= 36.3318 \\
\alpha^{(3)}_{1,1} &= 37.1514\\
\alpha^{(3)}_{1,2} &= 10.125
\end{aligned}\;,
\right.
\hspace{0.5cm}
\left\{ 
\begin{aligned}
\beta^{(3)}_{1, 0} &= -0.181866\\
\beta^{(3)}_{1, 1} &= 0.211083\\
\beta^{(3)}_{1, 2} &= -0.879515\\
\beta^{(3)}_{3, 0} &= -10.4385\\
\beta^{(3)}_{3, 2} &= 3.82702
\end{aligned}
\right.\;,
\hspace{0.5cm}
\left\{ 
\begin{aligned}
\beta^{(3)}_{5, 0} &= -70.9277\\
\beta^{(3)}_{5, 1} &= 56.3093 \\
\beta^{(3)}_{5, 2} &= 20.9951 \\
\beta^{(3)}_{5, 3} &= -7.55063
\end{aligned}
\right.\;.
\end{equation}

At least, one gives the error functions $\mathcal{E}^{(k)}$, 
\begin{align} \label{eq:ErrorFunctions}
\mathcal{E}^{(2)}(N^*_2,\omega) &= \begin{bmatrix} + 1 \\ 0\end{bmatrix} \sum_{n=N^*_2+1}^{\infty} \frac{\log^{1.5}n}{n^3}\, \omega^n\\
\mathcal{E}^{(3)}(N^*_3,\omega) &= \begin{bmatrix} + 15 \\ -15\end{bmatrix} \sum_{n=N^*_3+1}^{\infty} \frac{\log^{3}n}{n^2}\, \omega^n\;,
\end{align}
which encode the systematic error from the reconstructions. 

\subsection{Experimental data}

\begin{figure}[h]
\begin{center}
\input{TotalPlot}
\end{center}
\caption{Collection of the different experimental sets for the V-V spectrum. }
\label{fig:ExperimentalSpectrum2to11Gev}
\end{figure}
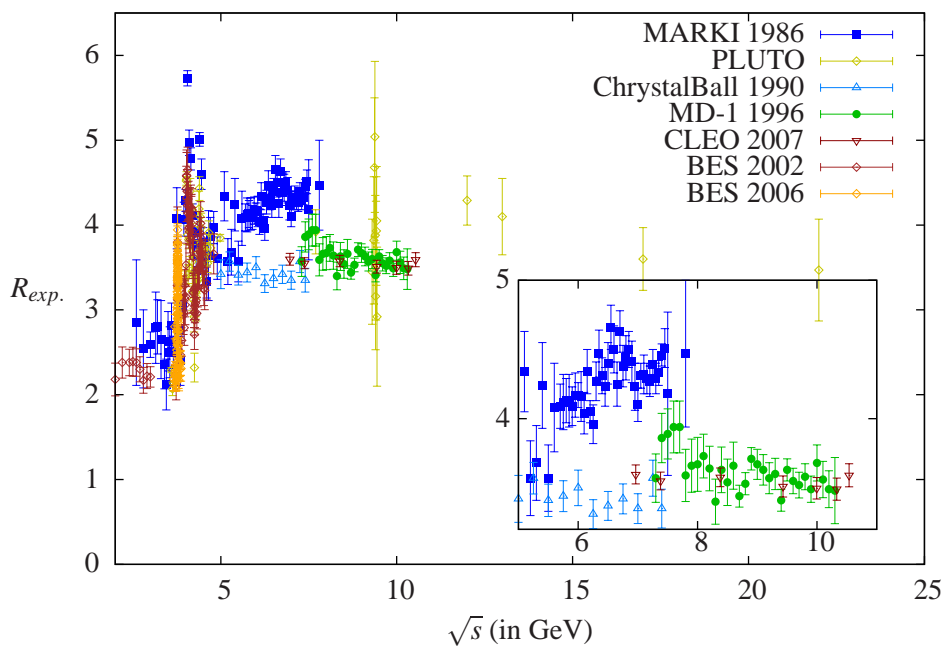

There exists several experimental results for the $e^+e^-$  in  hadrons that one can use for the fitting of the $c$ quark mass. Each of the experiments give the ratio $R(s)$ of the radiation-corrected measured hadronic cross section to the calculated lowest-order cross section for muon pair production, 
\begin{equation}\label{def:R}
R(s) = \frac{\sigma_0\left(e^ + e^ - \longrightarrow \text{hadrons} \right)}{\sigma_0\left(e^ + e^ - \longrightarrow \mu^ +\mu^ - \right)} = \frac{\sigma_0\left(e^ + e^ - \longrightarrow \text{hadrons} \right)}{4\pi \alpha^2/3s} \, ,
\end{equation}
that has the experimental values shown in Fig. \ref{fig:ExperimentalSpectrum2to11Gev} . 

\begin{table}[h]
\begin{center}
\begin{tabular}{|c|c|}
\hline\hline
Experiment			& Reference  \\ 
 \hline
MARK I 					&\cite{Siegrist:1981zp}\\
PLUTO					&\cite{Criegee:1981qx}\\
CrystalBall (Run 1)	&\cite{Edwards:1990pc}\\
CrystalBall (Run 2)	&\cite{Edwards:1990pc}\\
MD1						&\cite{Blinov:1993fw}\\
CLEO					&\cite{Ammar:1997sk}\\
CLEO 					&\cite{Besson:1984bd,Besson:2007aa}\\
BES						&\cite{Bai:2001ct}\\
BES						&\cite{Ablikim:2006mb}\\
CLEO					& \cite{:2007qwa}\\
CLEO					& \cite{CroninHennessy:2008yi}\\
\hline\hline
\end{tabular}
\caption{All different experimental sets considered for the fits.}
\label{tab:ReferencedSets}
\end{center}
\end{table}

This Fig \ref{fig:ExperimentalSpectrum2to11Gev} shows that the complete spectrum is sensitive to resonances, as expected. It is obvious that a perturbative approach cannot take into account the resonances description, then one has to make an arbitrary choice on where we assume that the continuum limit is reached or in other words, where the perturbative description is pertinent. Let's choose the value of 5 GeV. Of course the influence of the arbitrariness has to be discussed and taken account in the evaluation of the error but it is something depending on the perturbative and heavy-quark limit more than the reconstruction itself. 

The idea now is to perform a fit among all this data points to extract the \textit{perturbative} mass $m_c$ of the $c$-quark. 

\subsection{Fitting approach}

The first step in the fitting procedure is to choose the following expression for the running $\alpha_s(s)$, 
\begin{multline}
\alpha_s(s) =  \frac{4\pi}{\beta_0\ln(s/\Lambda^2)}\left[ 1 - \frac{2\beta_1}{\beta_0^2}\frac{\ln[\ln(s/\Lambda^2)]}{\ln(s/\Lambda^2)} \right.\\
\left.+\frac{4\beta_1^2}{\beta_0^4\ln^2(s/\Lambda^2)}\;\left(\left(\ln\left[\ln(s/\Lambda^2)\right]-\frac{1}{2}\right)^2 +\frac{\beta_2\beta_0}{8\beta_1^2}-\frac{5}{4}\right)\right] ,
\label{AlphaSrunning}
\end{multline}
where $\Lambda$ is the energy scale and the $\beta$-function has coefficients
\begin{align}
\beta_0 & =  11-\frac{2n_f}{3}\,, & \beta_1 & =  51-\frac{19n_f}{3}\,, & \beta_2 & = 2857-\frac{5033n_f}{9}+\frac{325n_f^2}{27}\;,
\end{align}
and $n_f$ is the number of quarks with mass smaller than $\sqrt{s}/2$.

The \textit{theoretical} expression (\ref{def:R}) is related to $\Pi(q^2)$ (\ref{Pi}), up to $\alpha_s^3$,  
\begin{multline}
\label{eq:RPi}
R_\text{th.}(s) = \left[\left(\frac{2}{3}\right)^2+\left(\frac{1}{3}\right)^2+\left(\frac{1}{3}\right)^2\right]N_c\left[ 1 + \frac{\alpha_s(s)}{\pi}+1.525 \left(\frac{\alpha_s(s)}{\pi}\right)^2-11.686\left(\frac{\alpha_s(s)}{\pi}\right)^3\right] \\ + 12\pi \left(\frac{2}{3}\right)^ 2 \I \left[\Pi^ {(0)} + \frac{4}{3} \frac{\alpha_s(s)}{\pi}\Pi^{(1)}+\left(\frac{\alpha_s(s)}{\pi}\right)^ 2\Pi^{(2)} + C_3 \left(\frac{\alpha_s(s)}{\pi}\right)^3 \Pi^{(3)}\right]
\end{multline}
where all $\Pi^{(k)}$ functions have the argument $z = \frac{s}{4m^2_c}$, and $N_c$ is the number of colors.

The goal of the analysis is to extract $m_c$ from the comparison between the value of $R_\text{exp.}$ and $R_\text{th.}$. The usual method used is to built the moments associated to R from 0 to $\Lambda^2$ and identifying the coefficients of the Taylor expansion that are proportional up to a factor to $m_c^{-2}$.  Instead of this approach, we propose to perform the analysis directly on the function itself, because thanks to the reconstruction method formula (\ref{approx}), its expression is available and its systematic error too (\ref{eq:ErrorFunctions}). 

For this we will use a $\chi^2$-method with the assumption  
\begin{equation}
\chi^ 2(m_c) \doteq \sum_{j=1}^ {N} \left(\frac{R_\text{exp.}(s_j)-R_\text{th.}(s_j)}{\sigma_\text{exp.}(s_j)}\right)^ 2 + \left(\frac{R_\text{exp.}(s_j)-R_\text{th.}(s_j)}{\sigma_\text{th.}(s_j)}\right)^ 2\;,
\end{equation}
where the $s_j$ are the experimental energy points, the $\sigma_\text{exp.}$ is the experimental error and the theoretical error  $\sigma_\text{th.}$ due the approximation of the reconstruction is given by 
\begin{multline}
\sigma_\text{th.}^2(s) = \frac{256\pi^2}{9} \left|\I \left[ \left(\frac{\alpha_s(s)}{\pi}\right)^ 2\mathcal{E}^{(2)}\left(N_2^*,\omega\right)\right]\right|^2 \\
+\frac{256\pi^2}{9} \; C_3^2 \; \left|\I \left[ \left(\frac{\alpha_s(s)}{\pi}\right)^ 3\mathcal{E}^{(3)}\left(N_3^*,\omega\right)\right]\right|^2\;,
\end{multline} 
with $\omega =\frac{1-\sqrt{1-\frac{s}{4m_c^2}}}{1+\sqrt{1-\frac{s}{4m_c^2}}}$.

\section{Results}

\subsection{Numerical results at order $\alpha_s^2$}

At $\alpha_s^ 2$ order, one obtains after a regression procedure with a $\chi^2_\text{min}/\text{d.o.f.}=1.03$, 
\begin{equation}
m_c(pole) = 1.85 \pm 0.08 \;\; \text{GeV}\;,
\end{equation}
that is translated into the $\overline{\text{MS}}$ mass as \cite{Melnikov:2000qh} 
\begin{equation}
m_c(\overline{\text{MS}}) = 1.12 \pm 0.08 \;\; \text{GeV}\;.
\end{equation}

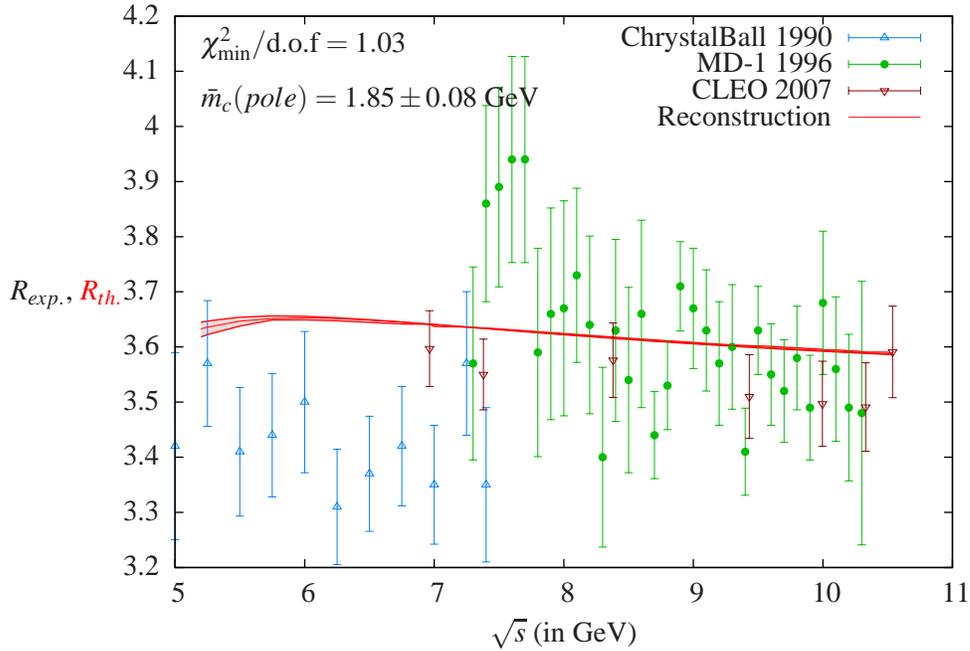
\begin{figure}[h]
\begin{center}
\input{ExtrapolationSTotal}
\end{center}
\caption{The reconstructed radiation-corrected measured hadronic cross section to the calculated lowest-order cross section for muon pair production.}
\label{fig:Extrapolation}
\end{figure}

 Assuming now that the mass $m_c$ obeys to a Gaussian density of probability, one can easily reconstruct points by points the error generated on $R_\text{th.}$ by this hypothesis, taking into account that the relation between $m_c$ and $R_\text{th.}$ is highly non linear and non trivial for expressing the error. We choose then to use a Monte-Carlo approach to  obtaining the mean value of $R_\text{th.}$ and its error as shown in Fig \ref{fig:Extrapolation}. 

\section{Conclusions} 

We show that it is possible to extract the charm mass value after a $\chi^2$ regression to the experimental data of the radiation-corrected measured hadronic cross section to the calculated lowest-order cross section for muon pair production using the non-analytic reconstruction of the heavy-quark correlators. We present here a preliminary result up to $\alpha_s^2$. The next step would include the order $\alpha_2^3$  and a complete analysis of all different systematic contributions \cite{GMM:2013}.

\end{document}

%% file: TotalPlot.tex
\begingroup%
\makeatletter%
\newcommand{\GNUPLOTspecial}{%
  \@sanitize\catcode`\%=14\relax\special}%
\setlength{\unitlength}{0.0500bp}%
\begin{picture}(7200,5040)(0,0)%
  {\GNUPLOTspecial{"
/gnudict 256 dict def
gnudict begin
%
%
/Color true def
/Blacktext true def
/Solid true def
/Dashlength 1 def
/Landscape false def
/Level1 false def
/Rounded false def
/ClipToBoundingBox false def
/TransparentPatterns false def
/gnulinewidth 5.000 def
/userlinewidth gnulinewidth def
/Gamma 1.0 def
/vshift -66 def
/dl1 {
  10.0 Dashlength mul mul
  Rounded { currentlinewidth 0.75 mul sub dup 0 le { pop 0.01 } if } if
} def
/dl2 {
  10.0 Dashlength mul mul
  Rounded { currentlinewidth 0.75 mul add } if
} def
/hpt_ 31.5 def
/vpt_ 31.5 def
/hpt hpt_ def
/vpt vpt_ def
Level1 {} {
/SDict 10 dict def
systemdict /pdfmark known not {
  userdict /pdfmark systemdict /cleartomark get put
} if
SDict begin [
  /Title (TotalPlot.tex)
  /Subject (gnuplot plot)
  /Creator (gnuplot 4.4 patchlevel 3)
  /Author (dg)
  /CreationDate (Mon Feb  4 09:55:43 2013)
  /DOCINFO pdfmark
end
} ifelse
/doclip {
  ClipToBoundingBox {
    newpath 0 0 moveto 360 0 lineto 360 252 lineto 0 252 lineto closepath
    clip
  } if
} def
%
%
%
/M {moveto} bind def
/L {lineto} bind def
/R {rmoveto} bind def
/V {rlineto} bind def
/N {newpath moveto} bind def
/Z {closepath} bind def
/C {setrgbcolor} bind def
/f {rlineto fill} bind def
/g {setgray} bind def
/Gshow {show} def   
/vpt2 vpt 2 mul def
/hpt2 hpt 2 mul def
/Lshow {currentpoint stroke M 0 vshift R 
	Blacktext {gsave 0 setgray show grestore} {show} ifelse} def
/Rshow {currentpoint stroke M dup stringwidth pop neg vshift R
	Blacktext {gsave 0 setgray show grestore} {show} ifelse} def
/Cshow {currentpoint stroke M dup stringwidth pop -2 div vshift R 
	Blacktext {gsave 0 setgray show grestore} {show} ifelse} def
/UP {dup vpt_ mul /vpt exch def hpt_ mul /hpt exch def
  /hpt2 hpt 2 mul def /vpt2 vpt 2 mul def} def
/DL {Color {setrgbcolor Solid {pop []} if 0 setdash}
 {pop pop pop 0 setgray Solid {pop []} if 0 setdash} ifelse} def
/BL {stroke userlinewidth 2 mul setlinewidth
	Rounded {1 setlinejoin 1 setlinecap} if} def
/AL {stroke userlinewidth 2 div setlinewidth
	Rounded {1 setlinejoin 1 setlinecap} if} def
/UL {dup gnulinewidth mul /userlinewidth exch def
	dup 1 lt {pop 1} if 10 mul /udl exch def} def
/PL {stroke userlinewidth setlinewidth
	Rounded {1 setlinejoin 1 setlinecap} if} def
3.8 setmiterlimit
/LCw {1 1 1} def
/LCb {0 0 0} def
/LCa {0 0 0} def
/LC0 {1 0 0} def
/LC1 {0 1 0} def
/LC2 {0 0 1} def
/LC3 {1 0 1} def
/LC4 {0 1 1} def
/LC5 {1 1 0} def
/LC6 {0 0 0} def
/LC7 {1 0.3 0} def
/LC8 {0.5 0.5 0.5} def
/LTw {PL [] 1 setgray} def
/LTb {BL [] LCb DL} def
/LTa {AL [1 udl mul 2 udl mul] 0 setdash LCa setrgbcolor} def
/LT0 {PL [] LC0 DL} def
/LT1 {PL [4 dl1 2 dl2] LC1 DL} def
/LT2 {PL [2 dl1 3 dl2] LC2 DL} def
/LT3 {PL [1 dl1 1.5 dl2] LC3 DL} def
/LT4 {PL [6 dl1 2 dl2 1 dl1 2 dl2] LC4 DL} def
/LT5 {PL [3 dl1 3 dl2 1 dl1 3 dl2] LC5 DL} def
/LT6 {PL [2 dl1 2 dl2 2 dl1 6 dl2] LC6 DL} def
/LT7 {PL [1 dl1 2 dl2 6 dl1 2 dl2 1 dl1 2 dl2] LC7 DL} def
/LT8 {PL [2 dl1 2 dl2 2 dl1 2 dl2 2 dl1 2 dl2 2 dl1 4 dl2] LC8 DL} def
/Pnt {stroke [] 0 setdash gsave 1 setlinecap M 0 0 V stroke grestore} def
/Dia {stroke [] 0 setdash 2 copy vpt add M
  hpt neg vpt neg V hpt vpt neg V
  hpt vpt V hpt neg vpt V closepath stroke
  Pnt} def
/Pls {stroke [] 0 setdash vpt sub M 0 vpt2 V
  currentpoint stroke M
  hpt neg vpt neg R hpt2 0 V stroke
 } def
/Box {stroke [] 0 setdash 2 copy exch hpt sub exch vpt add M
  0 vpt2 neg V hpt2 0 V 0 vpt2 V
  hpt2 neg 0 V closepath stroke
  Pnt} def
/Crs {stroke [] 0 setdash exch hpt sub exch vpt add M
  hpt2 vpt2 neg V currentpoint stroke M
  hpt2 neg 0 R hpt2 vpt2 V stroke} def
/TriU {stroke [] 0 setdash 2 copy vpt 1.12 mul add M
  hpt neg vpt -1.62 mul V
  hpt 2 mul 0 V
  hpt neg vpt 1.62 mul V closepath stroke
  Pnt} def
/Star {2 copy Pls Crs} def
/BoxF {stroke [] 0 setdash exch hpt sub exch vpt add M
  0 vpt2 neg V hpt2 0 V 0 vpt2 V
  hpt2 neg 0 V closepath fill} def
/TriUF {stroke [] 0 setdash vpt 1.12 mul add M
  hpt neg vpt -1.62 mul V
  hpt 2 mul 0 V
  hpt neg vpt 1.62 mul V closepath fill} def
/TriD {stroke [] 0 setdash 2 copy vpt 1.12 mul sub M
  hpt neg vpt 1.62 mul V
  hpt 2 mul 0 V
  hpt neg vpt -1.62 mul V closepath stroke
  Pnt} def
/TriDF {stroke [] 0 setdash vpt 1.12 mul sub M
  hpt neg vpt 1.62 mul V
  hpt 2 mul 0 V
  hpt neg vpt -1.62 mul V closepath fill} def
/DiaF {stroke [] 0 setdash vpt add M
  hpt neg vpt neg V hpt vpt neg V
  hpt vpt V hpt neg vpt V closepath fill} def
/Pent {stroke [] 0 setdash 2 copy gsave
  translate 0 hpt M 4 {72 rotate 0 hpt L} repeat
  closepath stroke grestore Pnt} def
/PentF {stroke [] 0 setdash gsave
  translate 0 hpt M 4 {72 rotate 0 hpt L} repeat
  closepath fill grestore} def
/Circle {stroke [] 0 setdash 2 copy
  hpt 0 360 arc stroke Pnt} def
/CircleF {stroke [] 0 setdash hpt 0 360 arc fill} def
/C0 {BL [] 0 setdash 2 copy moveto vpt 90 450 arc} bind def
/C1 {BL [] 0 setdash 2 copy moveto
	2 copy vpt 0 90 arc closepath fill
	vpt 0 360 arc closepath} bind def
/C2 {BL [] 0 setdash 2 copy moveto
	2 copy vpt 90 180 arc closepath fill
	vpt 0 360 arc closepath} bind def
/C3 {BL [] 0 setdash 2 copy moveto
	2 copy vpt 0 180 arc closepath fill
	vpt 0 360 arc closepath} bind def
/C4 {BL [] 0 setdash 2 copy moveto
	2 copy vpt 180 270 arc closepath fill
	vpt 0 360 arc closepath} bind def
/C5 {BL [] 0 setdash 2 copy moveto
	2 copy vpt 0 90 arc
	2 copy moveto
	2 copy vpt 180 270 arc closepath fill
	vpt 0 360 arc} bind def
/C6 {BL [] 0 setdash 2 copy moveto
	2 copy vpt 90 270 arc closepath fill
	vpt 0 360 arc closepath} bind def
/C7 {BL [] 0 setdash 2 copy moveto
	2 copy vpt 0 270 arc closepath fill
	vpt 0 360 arc closepath} bind def
/C8 {BL [] 0 setdash 2 copy moveto
	2 copy vpt 270 360 arc closepath fill
	vpt 0 360 arc closepath} bind def
/C9 {BL [] 0 setdash 2 copy moveto
	2 copy vpt 270 450 arc closepath fill
	vpt 0 360 arc closepath} bind def
/C10 {BL [] 0 setdash 2 copy 2 copy moveto vpt 270 360 arc closepath fill
	2 copy moveto
	2 copy vpt 90 180 arc closepath fill
	vpt 0 360 arc closepath} bind def
/C11 {BL [] 0 setdash 2 copy moveto
	2 copy vpt 0 180 arc closepath fill
	2 copy moveto
	2 copy vpt 270 360 arc closepath fill
	vpt 0 360 arc closepath} bind def
/C12 {BL [] 0 setdash 2 copy moveto
	2 copy vpt 180 360 arc closepath fill
	vpt 0 360 arc closepath} bind def
/C13 {BL [] 0 setdash 2 copy moveto
	2 copy vpt 0 90 arc closepath fill
	2 copy moveto
	2 copy vpt 180 360 arc closepath fill
	vpt 0 360 arc closepath} bind def
/C14 {BL [] 0 setdash 2 copy moveto
	2 copy vpt 90 360 arc closepath fill
	vpt 0 360 arc} bind def
/C15 {BL [] 0 setdash 2 copy vpt 0 360 arc closepath fill
	vpt 0 360 arc closepath} bind def
/Rec {newpath 4 2 roll moveto 1 index 0 rlineto 0 exch rlineto
	neg 0 rlineto closepath} bind def
/Square {dup Rec} bind def
/Bsquare {vpt sub exch vpt sub exch vpt2 Square} bind def
/S0 {BL [] 0 setdash 2 copy moveto 0 vpt rlineto BL Bsquare} bind def
/S1 {BL [] 0 setdash 2 copy vpt Square fill Bsquare} bind def
/S2 {BL [] 0 setdash 2 copy exch vpt sub exch vpt Square fill Bsquare} bind def
/S3 {BL [] 0 setdash 2 copy exch vpt sub exch vpt2 vpt Rec fill Bsquare} bind def
/S4 {BL [] 0 setdash 2 copy exch vpt sub exch vpt sub vpt Square fill Bsquare} bind def
/S5 {BL [] 0 setdash 2 copy 2 copy vpt Square fill
	exch vpt sub exch vpt sub vpt Square fill Bsquare} bind def
/S6 {BL [] 0 setdash 2 copy exch vpt sub exch vpt sub vpt vpt2 Rec fill Bsquare} bind def
/S7 {BL [] 0 setdash 2 copy exch vpt sub exch vpt sub vpt vpt2 Rec fill
	2 copy vpt Square fill Bsquare} bind def
/S8 {BL [] 0 setdash 2 copy vpt sub vpt Square fill Bsquare} bind def
/S9 {BL [] 0 setdash 2 copy vpt sub vpt vpt2 Rec fill Bsquare} bind def
/S10 {BL [] 0 setdash 2 copy vpt sub vpt Square fill 2 copy exch vpt sub exch vpt Square fill
	Bsquare} bind def
/S11 {BL [] 0 setdash 2 copy vpt sub vpt Square fill 2 copy exch vpt sub exch vpt2 vpt Rec fill
	Bsquare} bind def
/S12 {BL [] 0 setdash 2 copy exch vpt sub exch vpt sub vpt2 vpt Rec fill Bsquare} bind def
/S13 {BL [] 0 setdash 2 copy exch vpt sub exch vpt sub vpt2 vpt Rec fill
	2 copy vpt Square fill Bsquare} bind def
/S14 {BL [] 0 setdash 2 copy exch vpt sub exch vpt sub vpt2 vpt Rec fill
	2 copy exch vpt sub exch vpt Square fill Bsquare} bind def
/S15 {BL [] 0 setdash 2 copy Bsquare fill Bsquare} bind def
/D0 {gsave translate 45 rotate 0 0 S0 stroke grestore} bind def
/D1 {gsave translate 45 rotate 0 0 S1 stroke grestore} bind def
/D2 {gsave translate 45 rotate 0 0 S2 stroke grestore} bind def
/D3 {gsave translate 45 rotate 0 0 S3 stroke grestore} bind def
/D4 {gsave translate 45 rotate 0 0 S4 stroke grestore} bind def
/D5 {gsave translate 45 rotate 0 0 S5 stroke grestore} bind def
/D6 {gsave translate 45 rotate 0 0 S6 stroke grestore} bind def
/D7 {gsave translate 45 rotate 0 0 S7 stroke grestore} bind def
/D8 {gsave translate 45 rotate 0 0 S8 stroke grestore} bind def
/D9 {gsave translate 45 rotate 0 0 S9 stroke grestore} bind def
/D10 {gsave translate 45 rotate 0 0 S10 stroke grestore} bind def
/D11 {gsave translate 45 rotate 0 0 S11 stroke grestore} bind def
/D12 {gsave translate 45 rotate 0 0 S12 stroke grestore} bind def
/D13 {gsave translate 45 rotate 0 0 S13 stroke grestore} bind def
/D14 {gsave translate 45 rotate 0 0 S14 stroke grestore} bind def
/D15 {gsave translate 45 rotate 0 0 S15 stroke grestore} bind def
/DiaE {stroke [] 0 setdash vpt add M
  hpt neg vpt neg V hpt vpt neg V
  hpt vpt V hpt neg vpt V closepath stroke} def
/BoxE {stroke [] 0 setdash exch hpt sub exch vpt add M
  0 vpt2 neg V hpt2 0 V 0 vpt2 V
  hpt2 neg 0 V closepath stroke} def
/TriUE {stroke [] 0 setdash vpt 1.12 mul add M
  hpt neg vpt -1.62 mul V
  hpt 2 mul 0 V
  hpt neg vpt 1.62 mul V closepath stroke} def
/TriDE {stroke [] 0 setdash vpt 1.12 mul sub M
  hpt neg vpt 1.62 mul V
  hpt 2 mul 0 V
  hpt neg vpt -1.62 mul V closepath stroke} def
/PentE {stroke [] 0 setdash gsave
  translate 0 hpt M 4 {72 rotate 0 hpt L} repeat
  closepath stroke grestore} def
/CircE {stroke [] 0 setdash 
  hpt 0 360 arc stroke} def
/Opaque {gsave closepath 1 setgray fill grestore 0 setgray closepath} def
/DiaW {stroke [] 0 setdash vpt add M
  hpt neg vpt neg V hpt vpt neg V
  hpt vpt V hpt neg vpt V Opaque stroke} def
/BoxW {stroke [] 0 setdash exch hpt sub exch vpt add M
  0 vpt2 neg V hpt2 0 V 0 vpt2 V
  hpt2 neg 0 V Opaque stroke} def
/TriUW {stroke [] 0 setdash vpt 1.12 mul add M
  hpt neg vpt -1.62 mul V
  hpt 2 mul 0 V
  hpt neg vpt 1.62 mul V Opaque stroke} def
/TriDW {stroke [] 0 setdash vpt 1.12 mul sub M
  hpt neg vpt 1.62 mul V
  hpt 2 mul 0 V
  hpt neg vpt -1.62 mul V Opaque stroke} def
/PentW {stroke [] 0 setdash gsave
  translate 0 hpt M 4 {72 rotate 0 hpt L} repeat
  Opaque stroke grestore} def
/CircW {stroke [] 0 setdash 
  hpt 0 360 arc Opaque stroke} def
/BoxFill {gsave Rec 1 setgray fill grestore} def
/Density {
  /Fillden exch def
  currentrgbcolor
  /ColB exch def /ColG exch def /ColR exch def
  /ColR ColR Fillden mul Fillden sub 1 add def
  /ColG ColG Fillden mul Fillden sub 1 add def
  /ColB ColB Fillden mul Fillden sub 1 add def
  ColR ColG ColB setrgbcolor} def
/BoxColFill {gsave Rec PolyFill} def
/PolyFill {gsave Density fill grestore grestore} def
/h {rlineto rlineto rlineto gsave closepath fill grestore} bind def
%
%
/PatternFill {gsave /PFa [ 9 2 roll ] def
  PFa 0 get PFa 2 get 2 div add PFa 1 get PFa 3 get 2 div add translate
  PFa 2 get -2 div PFa 3 get -2 div PFa 2 get PFa 3 get Rec
  gsave 1 setgray fill grestore clip
  currentlinewidth 0.5 mul setlinewidth
  /PFs PFa 2 get dup mul PFa 3 get dup mul add sqrt def
  0 0 M PFa 5 get rotate PFs -2 div dup translate
  0 1 PFs PFa 4 get div 1 add floor cvi
	{PFa 4 get mul 0 M 0 PFs V} for
  0 PFa 6 get ne {
	0 1 PFs PFa 4 get div 1 add floor cvi
	{PFa 4 get mul 0 2 1 roll M PFs 0 V} for
 } if
  stroke grestore} def
/languagelevel where
 {pop languagelevel} {1} ifelse
 2 lt
	{/InterpretLevel1 true def}
	{/InterpretLevel1 Level1 def}
 ifelse
%
%
/Level2PatternFill {
/Tile8x8 {/PaintType 2 /PatternType 1 /TilingType 1 /BBox [0 0 8 8] /XStep 8 /YStep 8}
	bind def
/KeepColor {currentrgbcolor [/Pattern /DeviceRGB] setcolorspace} bind def
<< Tile8x8
 /PaintProc {0.5 setlinewidth pop 0 0 M 8 8 L 0 8 M 8 0 L stroke} 
>> matrix makepattern
/Pat1 exch def
<< Tile8x8
 /PaintProc {0.5 setlinewidth pop 0 0 M 8 8 L 0 8 M 8 0 L stroke
	0 4 M 4 8 L 8 4 L 4 0 L 0 4 L stroke}
>> matrix makepattern
/Pat2 exch def
<< Tile8x8
 /PaintProc {0.5 setlinewidth pop 0 0 M 0 8 L
	8 8 L 8 0 L 0 0 L fill}
>> matrix makepattern
/Pat3 exch def
<< Tile8x8
 /PaintProc {0.5 setlinewidth pop -4 8 M 8 -4 L
	0 12 M 12 0 L stroke}
>> matrix makepattern
/Pat4 exch def
<< Tile8x8
 /PaintProc {0.5 setlinewidth pop -4 0 M 8 12 L
	0 -4 M 12 8 L stroke}
>> matrix makepattern
/Pat5 exch def
<< Tile8x8
 /PaintProc {0.5 setlinewidth pop -2 8 M 4 -4 L
	0 12 M 8 -4 L 4 12 M 10 0 L stroke}
>> matrix makepattern
/Pat6 exch def
<< Tile8x8
 /PaintProc {0.5 setlinewidth pop -2 0 M 4 12 L
	0 -4 M 8 12 L 4 -4 M 10 8 L stroke}
>> matrix makepattern
/Pat7 exch def
<< Tile8x8
 /PaintProc {0.5 setlinewidth pop 8 -2 M -4 4 L
	12 0 M -4 8 L 12 4 M 0 10 L stroke}
>> matrix makepattern
/Pat8 exch def
<< Tile8x8
 /PaintProc {0.5 setlinewidth pop 0 -2 M 12 4 L
	-4 0 M 12 8 L -4 4 M 8 10 L stroke}
>> matrix makepattern
/Pat9 exch def
/Pattern1 {PatternBgnd KeepColor Pat1 setpattern} bind def
/Pattern2 {PatternBgnd KeepColor Pat2 setpattern} bind def
/Pattern3 {PatternBgnd KeepColor Pat3 setpattern} bind def
/Pattern4 {PatternBgnd KeepColor Landscape {Pat5} {Pat4} ifelse setpattern} bind def
/Pattern5 {PatternBgnd KeepColor Landscape {Pat4} {Pat5} ifelse setpattern} bind def
/Pattern6 {PatternBgnd KeepColor Landscape {Pat9} {Pat6} ifelse setpattern} bind def
/Pattern7 {PatternBgnd KeepColor Landscape {Pat8} {Pat7} ifelse setpattern} bind def
} def
%
%
%
/PatternBgnd {
  TransparentPatterns {} {gsave 1 setgray fill grestore} ifelse
} def
%
%
/Level1PatternFill {
/Pattern1 {0.250 Density} bind def
/Pattern2 {0.500 Density} bind def
/Pattern3 {0.750 Density} bind def
/Pattern4 {0.125 Density} bind def
/Pattern5 {0.375 Density} bind def
/Pattern6 {0.625 Density} bind def
/Pattern7 {0.875 Density} bind def
} def
%
%
Level1 {Level1PatternFill} {Level2PatternFill} ifelse
/Symbol-Oblique /Symbol findfont [1 0 .167 1 0 0] makefont
dup length dict begin {1 index /FID eq {pop pop} {def} ifelse} forall
currentdict end definefont pop
end
gnudict begin
gsave
doclip
0 0 translate
0.050 0.050 scale
0 setgray
newpath
1.000 UL
LTb
740 640 M
63 0 V
6036 0 R
-63 0 V
740 1280 M
63 0 V
6036 0 R
-63 0 V
740 1920 M
63 0 V
6036 0 R
-63 0 V
740 2560 M
63 0 V
6036 0 R
-63 0 V
740 3199 M
63 0 V
6036 0 R
-63 0 V
740 3839 M
63 0 V
6036 0 R
-63 0 V
740 4479 M
63 0 V
6036 0 R
-63 0 V
1536 640 M
0 63 V
0 4096 R
0 -63 V
2861 640 M
0 63 V
0 4096 R
0 -63 V
4187 640 M
0 63 V
0 4096 R
0 -63 V
5513 640 M
0 63 V
0 4096 R
0 -63 V
6839 640 M
0 63 V
0 4096 R
0 -63 V
stroke
740 4799 N
740 640 L
6099 0 V
0 4159 V
-6099 0 V
Z stroke
LCb setrgbcolor
LTb
LCb setrgbcolor
LTb
1.000 UP
1.000 UL
LTb
1.000 UP
1.000 UL
LT0
0.00 0.00 1.00 C LCb setrgbcolor
LT0
0.00 0.00 1.00 C 6056 4636 M
543 0 V
-543 31 R
0 -62 V
543 62 R
0 -62 V
899 1990 M
0 947 V
868 1990 M
62 0 V
-62 947 R
62 0 V
22 -966 R
0 589 V
921 1971 M
62 0 V
-62 589 R
62 0 V
22 -359 R
0 192 V
974 2201 M
62 0 V
-62 192 R
62 0 V
9 -237 R
0 538 V
-31 -538 R
62 0 V
-62 538 R
62 0 V
-18 -467 R
0 409 V
-31 -409 R
62 0 V
-62 409 R
62 0 V
-4 -595 R
0 589 V
-31 -589 R
62 0 V
-62 589 R
62 0 V
-5 -666 R
0 359 V
-31 -359 R
62 0 V
-62 359 R
62 0 V
-17 -518 R
0 383 V
-31 -383 R
62 0 V
-62 383 R
62 0 V
-18 -89 R
0 448 V
-31 -448 R
62 0 V
-62 448 R
62 0 V
-18 -499 R
0 384 V
-31 -384 R
62 0 V
-62 384 R
62 0 V
-18 -154 R
0 333 V
-31 -333 R
62 0 V
-62 333 R
62 0 V
-17 -493 R
0 243 V
-31 -243 R
62 0 V
-62 243 R
62 0 V
-5 659 R
0 461 V
-31 -461 R
62 0 V
-62 461 R
62 0 V
1217 2240 M
0 332 V
-31 -332 R
62 0 V
-62 332 R
62 0 V
-17 -582 R
0 371 V
-31 -371 R
62 0 V
-62 371 R
62 0 V
-18 141 R
0 166 V
-31 -166 R
62 0 V
-62 166 R
62 0 V
-18 448 R
0 256 V
-31 -256 R
62 0 V
stroke 1288 3116 M
-62 256 R
62 0 V
-18 -70 R
0 166 V
-31 -166 R
62 0 V
-62 166 R
62 0 V
-17 781 R
0 115 V
-31 -115 R
62 0 V
-62 115 R
62 0 V
-18 -640 R
0 192 V
-31 -192 R
62 0 V
-62 192 R
62 0 V
-18 -301 R
0 167 V
-31 -167 R
62 0 V
-62 167 R
62 0 V
-18 -602 R
0 179 V
-31 -179 R
62 0 V
-62 179 R
62 0 V
-17 -416 R
0 231 V
-31 -231 R
62 0 V
-62 231 R
62 0 V
-18 -410 R
0 192 V
-31 -192 R
62 0 V
-62 192 R
62 0 V
-18 64 R
0 243 V
-31 -243 R
62 0 V
-62 243 R
62 0 V
-18 531 R
0 103 V
-31 -103 R
62 0 V
-62 103 R
62 0 V
-17 -429 R
0 230 V
-31 -230 R
62 0 V
-62 230 R
62 0 V
-18 -748 R
0 230 V
-31 -230 R
62 0 V
-62 230 R
62 0 V
-18 -429 R
0 320 V
-31 -320 R
62 0 V
-62 320 R
62 0 V
-18 -422 R
0 243 V
-31 -243 R
62 0 V
-62 243 R
62 0 V
-17 -262 R
0 678 V
-31 -678 R
62 0 V
-62 678 R
62 0 V
-18 -345 R
0 294 V
-31 -294 R
62 0 V
-62 294 R
62 0 V
-5 -198 R
0 243 V
-31 -243 R
62 0 V
-62 243 R
62 0 V
-4 -512 R
0 320 V
-31 -320 R
62 0 V
-62 320 R
62 0 V
stroke 1540 3110 M
22 121 R
0 371 V
-31 -371 R
62 0 V
-62 371 R
62 0 V
-4 -851 R
0 346 V
-31 -346 R
62 0 V
-62 346 R
62 0 V
-5 -275 R
0 345 V
-31 -345 R
62 0 V
-62 345 R
62 0 V
-4 -12 R
0 396 V
-31 -396 R
62 0 V
-62 396 R
62 0 V
-5 -780 R
0 307 V
-31 -307 R
62 0 V
-62 307 R
62 0 V
-4 -32 R
0 409 V
-31 -409 R
62 0 V
-62 409 R
62 0 V
-5 -300 R
0 204 V
-31 -204 R
62 0 V
-62 204 R
62 0 V
-18 -211 R
0 256 V
-31 -256 R
62 0 V
-62 256 R
62 0 V
-17 -224 R
0 205 V
-31 -205 R
62 0 V
-62 205 R
62 0 V
-18 -224 R
0 243 V
-31 -243 R
62 0 V
-62 243 R
62 0 V
-18 -237 R
0 180 V
-31 -180 R
62 0 V
-62 180 R
62 0 V
-18 -141 R
0 205 V
-31 -205 R
62 0 V
-62 205 R
62 0 V
-17 -160 R
0 115 V
-31 -115 R
62 0 V
-62 115 R
62 0 V
-18 -179 R
0 230 V
-31 -230 R
62 0 V
-62 230 R
62 0 V
-18 -288 R
0 192 V
-31 -192 R
62 0 V
-62 192 R
62 0 V
-18 -6 R
0 204 V
-31 -204 R
62 0 V
-62 204 R
62 0 V
-17 -339 R
0 103 V
-31 -103 R
62 0 V
-62 103 R
62 0 V
-18 -199 R
0 179 V
stroke 1867 3263 M
-31 -179 R
62 0 V
-62 179 R
62 0 V
-18 20 R
0 179 V
-31 -179 R
62 0 V
-62 179 R
62 0 V
-17 -71 R
0 218 V
-31 -218 R
62 0 V
-62 218 R
62 0 V
-18 -294 R
0 166 V
-31 -166 R
62 0 V
-62 166 R
62 0 V
-18 -224 R
0 179 V
-31 -179 R
62 0 V
-62 179 R
62 0 V
-18 -77 R
0 192 V
-31 -192 R
62 0 V
-62 192 R
62 0 V
-17 -32 R
0 205 V
-31 -205 R
62 0 V
-62 205 R
62 0 V
-18 -313 R
0 217 V
-31 -217 R
62 0 V
-62 217 R
62 0 V
-18 -371 R
0 205 V
-31 -205 R
62 0 V
-62 205 R
62 0 V
-18 45 R
0 191 V
-31 -191 R
62 0 V
-62 191 R
62 0 V
-17 -351 R
0 192 V
-31 -192 R
62 0 V
-62 192 R
62 0 V
-18 -160 R
0 204 V
-31 -204 R
62 0 V
-62 204 R
62 0 V
-18 -147 R
0 166 V
-31 -166 R
62 0 V
-62 166 R
62 0 V
-18 -236 R
0 192 V
-31 -192 R
62 0 V
-62 192 R
62 0 V
-17 -320 R
0 217 V
-31 -217 R
62 0 V
-62 217 R
62 0 V
-18 -268 R
0 153 V
-31 -153 R
62 0 V
-62 153 R
62 0 V
-18 0 R
0 115 V
-31 -115 R
62 0 V
-62 115 R
62 0 V
-18 -140 R
0 179 V
-31 -179 R
62 0 V
stroke 2123 3315 M
-62 179 R
62 0 V
-17 -179 R
0 140 V
-31 -140 R
62 0 V
-62 140 R
62 0 V
-18 -153 R
0 141 V
-31 -141 R
62 0 V
-62 141 R
62 0 V
-18 -64 R
0 140 V
-31 -140 R
62 0 V
-62 140 R
62 0 V
-18 -204 R
0 140 V
-31 -140 R
62 0 V
-62 140 R
62 0 V
-17 -102 R
0 115 V
-31 -115 R
62 0 V
-62 115 R
62 0 V
-18 -25 R
0 102 V
-31 -102 R
62 0 V
-62 102 R
62 0 V
-18 -109 R
0 179 V
-31 -179 R
62 0 V
-62 179 R
62 0 V
-18 -678 R
0 755 V
-31 -755 R
62 0 V
-62 755 R
62 0 V
49 -531 R
0 678 V
-31 -678 R
62 0 V
-62 678 R
62 0 V
899 2464 BoxF
952 2265 BoxF
1005 2297 BoxF
1045 2425 BoxF
1058 2432 BoxF
1085 2336 BoxF
1111 2144 BoxF
1125 1996 BoxF
1138 2323 BoxF
1151 2240 BoxF
1164 2444 BoxF
1178 2240 BoxF
1204 3251 BoxF
1217 2406 BoxF
1231 2176 BoxF
1244 2585 BoxF
1257 3244 BoxF
1270 3385 BoxF
1284 4306 BoxF
1297 3820 BoxF
1310 3698 BoxF
1323 3270 BoxF
1337 3059 BoxF
1350 2860 BoxF
1363 3142 BoxF
1376 3846 BoxF
1390 3583 BoxF
1403 3065 BoxF
1416 2911 BoxF
1429 2771 BoxF
1443 2969 BoxF
1456 3110 BoxF
1482 3180 BoxF
1509 2950 BoxF
1562 3417 BoxF
1589 2924 BoxF
1615 2995 BoxF
1642 3353 BoxF
1668 2924 BoxF
1695 3251 BoxF
1721 3257 BoxF
1734 3276 BoxF
1748 3283 BoxF
1761 3283 BoxF
1774 3257 BoxF
1787 3308 BoxF
1801 3308 BoxF
1814 3302 BoxF
1827 3225 BoxF
1840 3417 BoxF
1854 3231 BoxF
1867 3174 BoxF
1880 3372 BoxF
1894 3500 BoxF
1907 3398 BoxF
1920 3347 BoxF
1933 3455 BoxF
1947 3622 BoxF
1960 3519 BoxF
1973 3359 BoxF
1986 3602 BoxF
2000 3443 BoxF
2013 3481 BoxF
2026 3519 BoxF
2039 3462 BoxF
2053 3347 BoxF
2066 3263 BoxF
2079 3398 BoxF
2092 3404 BoxF
2106 3385 BoxF
2119 3372 BoxF
2132 3449 BoxF
2145 3385 BoxF
2159 3411 BoxF
2172 3494 BoxF
2185 3526 BoxF
2198 3315 BoxF
2278 3500 BoxF
6327 4636 BoxF
1.000 UP
1.000 UL
LT1
0.78 0.78 0.00 C LCb setrgbcolor
LT1
0.78 0.78 0.00 C 6056 4436 M
543 0 V
-543 31 R
0 -62 V
543 62 R
0 -62 V
1164 2086 M
0 38 V
-31 -38 R
62 0 V
-62 38 R
62 0 V
-23 -211 R
0 103 V
-31 -103 R
62 0 V
-62 103 R
62 0 V
-23 -71 R
0 115 V
-31 -115 R
62 0 V
-62 115 R
62 0 V
59 691 R
0 103 V
-31 -103 R
62 0 V
-62 103 R
62 0 V
-25 646 R
0 64 V
-31 -64 R
62 0 V
-62 64 R
62 0 V
-29 -77 R
0 115 V
-31 -115 R
62 0 V
-62 115 R
62 0 V
-25 -550 R
0 141 V
-31 -141 R
62 0 V
-62 141 R
62 0 V
-26 -435 R
0 141 V
-31 -141 R
62 0 V
-62 141 R
62 0 V
-26 121 R
0 141 V
-31 -141 R
62 0 V
-62 141 R
62 0 V
-25 -365 R
0 167 V
-31 -167 R
62 0 V
-62 167 R
62 0 V
-26 396 R
0 192 V
-31 -192 R
62 0 V
-62 192 R
62 0 V
-26 -339 R
0 218 V
-31 -218 R
62 0 V
-62 218 R
62 0 V
-23 -634 R
0 256 V
-31 -256 R
62 0 V
-62 256 R
62 0 V
-28 -499 R
0 294 V
-31 -294 R
62 0 V
-62 294 R
62 0 V
-23 -313 R
0 205 V
-31 -205 R
62 0 V
-62 205 R
62 0 V
-26 -346 R
0 231 V
-31 -231 R
62 0 V
-62 231 R
62 0 V
-28 -608 R
0 217 V
-31 -217 R
62 0 V
stroke 1368 2016 M
-62 217 R
62 0 V
-23 275 R
0 371 V
-31 -371 R
62 0 V
-62 371 R
62 0 V
-26 -403 R
0 359 V
-31 -359 R
62 0 V
-62 359 R
62 0 V
-28 -154 R
0 192 V
-31 -192 R
62 0 V
-62 192 R
62 0 V
-26 6 R
0 205 V
-31 -205 R
62 0 V
-62 205 R
62 0 V
-26 -384 R
0 167 V
-31 -167 R
62 0 V
-62 167 R
62 0 V
-23 320 R
0 268 V
-31 -268 R
62 0 V
-62 268 R
62 0 V
-28 -70 R
0 179 V
-31 -179 R
62 0 V
-62 179 R
62 0 V
-26 -339 R
0 77 V
-31 -77 R
62 0 V
-62 77 R
62 0 V
-23 -231 R
0 218 V
-31 -218 R
62 0 V
-62 218 R
62 0 V
-26 -486 R
0 204 V
-31 -204 R
62 0 V
-62 204 R
62 0 V
-20 -32 R
0 90 V
-31 -90 R
62 0 V
-62 90 R
62 0 V
-20 -90 R
0 90 V
-31 -90 R
62 0 V
-62 90 R
62 0 V
-24 -166 R
0 166 V
-31 -166 R
62 0 V
-62 166 R
62 0 V
-17 -224 R
0 141 V
-31 -141 R
62 0 V
-62 141 R
62 0 V
-21 70 R
0 179 V
-31 -179 R
62 0 V
-62 179 R
62 0 V
-17 -307 R
0 205 V
-31 -205 R
62 0 V
-62 205 R
62 0 V
40 -51 R
0 51 V
-31 -51 R
62 0 V
-62 51 R
62 0 V
stroke 1561 3129 M
690 -147 R
0 333 V
-31 -333 R
62 0 V
-62 333 R
62 0 V
394 -864 R
0 793 V
-31 -793 R
62 0 V
-62 793 R
62 0 V
-18 -640 R
0 960 V
-31 -960 R
62 0 V
-62 960 R
62 0 V
-26 -454 R
0 1049 V
2663 3110 M
62 0 V
-62 1049 R
62 0 V
-28 -864 R
0 1139 V
2666 3295 M
62 0 V
-62 1139 R
62 0 V
2700 2822 M
0 601 V
-31 -601 R
62 0 V
-62 601 R
62 0 V
2702 2259 M
0 806 V
-31 -806 R
62 0 V
-62 806 R
62 0 V
-28 -582 R
0 896 V
-31 -896 R
62 0 V
-62 896 R
62 0 V
-28 -634 R
0 819 V
-31 -819 R
62 0 V
-62 819 R
62 0 V
-29 -742 R
0 819 V
-31 -819 R
62 0 V
-62 819 R
62 0 V
2713 1984 M
0 1049 V
2682 1984 M
62 0 V
-62 1049 R
62 0 V
648 166 R
0 371 V
-31 -371 R
62 0 V
-62 371 R
62 0 V
234 -595 R
0 576 V
-31 -576 R
62 0 V
-62 576 R
62 0 V
4718 2707 M
0 473 V
-31 -473 R
62 0 V
-62 473 R
62 0 V
6043 2476 M
0 768 V
-31 -768 R
62 0 V
-62 768 R
62 0 V
1164 2105 Dia
1172 1964 Dia
1180 2003 Dia
1270 2803 Dia
1276 3532 Dia
1278 3545 Dia
1284 3123 Dia
1289 2828 Dia
1294 3091 Dia
1300 2879 Dia
1305 3455 Dia
1310 3321 Dia
1318 2924 Dia
1321 2700 Dia
1329 2636 Dia
1334 2508 Dia
1337 2124 Dia
1345 2694 Dia
1350 2656 Dia
1353 2777 Dia
1358 2982 Dia
1363 2783 Dia
1371 3321 Dia
1374 3475 Dia
1379 3263 Dia
1387 3180 Dia
1392 2905 Dia
1403 3020 Dia
1414 3020 Dia
1421 2982 Dia
1435 2911 Dia
1445 3142 Dia
1459 3027 Dia
1530 3103 Dia
2251 3148 Dia
2676 2847 Dia
2689 3084 Dia
2694 3634 Dia
2697 3865 Dia
2700 3123 Dia
2702 2662 Dia
2705 2931 Dia
2708 3155 Dia
2710 3231 Dia
2713 2508 Dia
3392 3385 Dia
3657 3263 Dia
4718 2943 Dia
6043 2860 Dia
6327 4436 Dia
1.000 UP
1.000 UL
LT2
0.00 0.50 1.00 C LCb setrgbcolor
LT2
0.00 0.50 1.00 C 6056 4236 M
543 0 V
-543 31 R
0 -62 V
543 62 R
0 -62 V
1536 2720 M
0 217 V
-31 -217 R
62 0 V
-62 217 R
62 0 V
35 -86 R
0 146 V
-31 -146 R
62 0 V
-62 146 R
62 0 V
35 -250 R
0 149 V
-31 -149 R
62 0 V
-62 149 R
62 0 V
35 -126 R
0 143 V
-31 -143 R
62 0 V
-62 143 R
62 0 V
36 -115 R
0 163 V
-31 -163 R
62 0 V
-62 163 R
62 0 V
35 -270 R
0 134 V
-31 -134 R
62 0 V
-62 134 R
62 0 V
35 -96 R
0 134 V
-31 -134 R
62 0 V
-62 134 R
62 0 V
36 -104 R
0 138 V
-31 -138 R
62 0 V
-62 138 R
62 0 V
35 -182 R
0 137 V
-31 -137 R
62 0 V
-62 137 R
62 0 V
35 -11 R
0 167 V
-31 -167 R
62 0 V
-62 167 R
62 0 V
9 -314 R
0 179 V
-31 -179 R
62 0 V
-62 179 R
62 0 V
1536 2828 TriU
1602 2924 TriU
1668 2822 TriU
1734 2841 TriU
1801 2879 TriU
1867 2758 TriU
1933 2796 TriU
2000 2828 TriU
2066 2783 TriU
2132 2924 TriU
2172 2783 TriU
6327 4236 TriU
1.000 UP
1.000 UL
LT3
0.00 0.75 0.00 C LCb setrgbcolor
LT3
0.00 0.75 0.00 C 6056 4036 M
543 0 V
-543 31 R
0 -62 V
543 62 R
0 -62 V
2145 2812 M
0 224 V
-31 -224 R
62 0 V
-62 224 R
62 0 V
-4 -40 R
0 228 V
-31 -228 R
62 0 V
-62 228 R
62 0 V
-5 -211 R
0 232 V
-31 -232 R
62 0 V
-62 232 R
62 0 V
-4 -204 R
0 240 V
-31 -240 R
62 0 V
-62 240 R
62 0 V
-5 -240 R
0 240 V
-31 -240 R
62 0 V
-62 240 R
62 0 V
-4 -465 R
0 242 V
-31 -242 R
62 0 V
-62 242 R
62 0 V
-4 -199 R
0 246 V
-31 -246 R
62 0 V
-62 246 R
62 0 V
-5 -242 R
0 250 V
-31 -250 R
62 0 V
-62 250 R
62 0 V
-4 -187 R
0 202 V
-31 -202 R
62 0 V
-62 202 R
62 0 V
-5 -262 R
0 206 V
-31 -206 R
62 0 V
-62 206 R
62 0 V
-4 -361 R
0 209 V
-31 -209 R
62 0 V
-62 209 R
62 0 V
-5 -63 R
0 211 V
-31 -211 R
62 0 V
-62 211 R
62 0 V
-4 -270 R
0 215 V
-31 -215 R
62 0 V
-62 215 R
62 0 V
-5 -140 R
0 218 V
-31 -218 R
62 0 V
-62 218 R
62 0 V
-4 -300 R
0 101 V
-31 -101 R
62 0 V
-62 101 R
62 0 V
-5 -45 R
0 103 V
-31 -103 R
62 0 V
-62 103 R
62 0 V
-4 12 R
0 104 V
-31 -104 R
62 0 V
stroke 2601 2962 M
-62 104 R
62 0 V
-5 -148 R
0 140 V
-31 -140 R
62 0 V
-62 140 R
62 0 V
-4 -166 R
0 141 V
-31 -141 R
62 0 V
-62 141 R
62 0 V
-5 -180 R
0 143 V
-31 -143 R
62 0 V
-62 143 R
62 0 V
-4 -125 R
0 145 V
-31 -145 R
62 0 V
-62 145 R
62 0 V
-5 -245 R
0 101 V
-31 -101 R
62 0 V
-62 101 R
62 0 V
-4 39 R
0 103 V
-31 -103 R
62 0 V
-62 103 R
62 0 V
-5 -161 R
0 117 V
-31 -117 R
62 0 V
-62 117 R
62 0 V
-4 -137 R
0 119 V
-31 -119 R
62 0 V
-62 119 R
62 0 V
-5 -81 R
0 120 V
-31 -120 R
62 0 V
-62 120 R
62 0 V
-4 -179 R
0 122 V
-31 -122 R
62 0 V
-62 122 R
62 0 V
-5 -23 R
0 167 V
-31 -167 R
62 0 V
-62 167 R
62 0 V
-4 -244 R
0 168 V
-31 -168 R
62 0 V
-62 168 R
62 0 V
-5 -214 R
0 170 V
-31 -170 R
62 0 V
-62 170 R
62 0 V
-4 -244 R
0 306 V
-31 -306 R
62 0 V
-62 306 R
62 0 V
2145 2924 CircleF
2172 3110 CircleF
2198 3129 CircleF
2225 3161 CircleF
2251 3161 CircleF
2278 2937 CircleF
2305 2982 CircleF
2331 2988 CircleF
2358 3027 CircleF
2384 2969 CircleF
2411 2815 CircleF
2437 2963 CircleF
2464 2905 CircleF
2490 2982 CircleF
2517 2841 CircleF
2543 2899 CircleF
2570 3014 CircleF
2596 2988 CircleF
2623 2963 CircleF
2649 2924 CircleF
2676 2943 CircleF
2702 2822 CircleF
2729 2963 CircleF
2755 2911 CircleF
2782 2892 CircleF
2808 2931 CircleF
2835 2873 CircleF
2861 2995 CircleF
2888 2918 CircleF
2914 2873 CircleF
2941 2867 CircleF
6327 4036 CircleF
1.000 UP
1.000 UL
LT4
0.55 0.00 0.00 C LCb setrgbcolor
LT4
0.55 0.00 0.00 C 6056 3836 M
543 0 V
-543 31 R
0 -62 V
543 62 R
0 -62 V
3004 2885 M
0 106 V
-31 -106 R
62 0 V
-62 106 R
62 0 V
-86 -169 R
0 103 V
-31 -103 R
62 0 V
-62 103 R
62 0 V
-120 -97 R
0 99 V
-31 -99 R
62 0 V
-62 99 R
62 0 V
-180 -90 R
0 97 V
-31 -97 R
62 0 V
-62 97 R
62 0 V
-310 -49 R
0 86 V
-31 -86 R
62 0 V
-62 86 R
62 0 V
2167 2870 M
0 83 V
-31 -83 R
62 0 V
-62 83 R
62 0 V
-142 -56 R
0 89 V
-31 -89 R
62 0 V
-62 89 R
62 0 V
3004 2938 TriD
2949 2874 TriD
2860 2878 TriD
2711 2886 TriD
2432 2928 TriD
2167 2911 TriD
2056 2942 TriD
6327 3836 TriD
1.000 UP
1.000 UL
LT5
0.65 0.16 0.16 C LCb setrgbcolor
LT5
0.65 0.16 0.16 C 6056 3636 M
543 0 V
-543 31 R
0 -62 V
543 62 R
0 -62 V
740 1911 M
0 247 V
709 1911 M
62 0 V
-62 247 R
62 0 V
470 52 R
0 238 V
-31 -238 R
62 0 V
-62 238 R
62 0 V
30 609 R
0 425 V
-31 -425 R
62 0 V
-62 425 R
62 0 V
28 -900 R
0 300 V
-31 -300 R
62 0 V
-62 300 R
62 0 V
793 2045 M
0 235 V
762 2045 M
62 0 V
-62 235 R
62 0 V
428 254 R
0 282 V
-31 -282 R
62 0 V
-62 282 R
62 0 V
22 232 R
0 290 V
-31 -290 R
62 0 V
-62 290 R
62 0 V
27 -592 R
0 254 V
-31 -254 R
62 0 V
-62 254 R
62 0 V
846 2063 M
0 200 V
815 2063 M
62 0 V
-62 200 R
62 0 V
377 111 R
0 294 V
-31 -294 R
62 0 V
-62 294 R
62 0 V
22 273 R
0 300 V
-31 -300 R
62 0 V
-62 300 R
62 0 V
28 -523 R
0 284 V
-31 -284 R
62 0 V
-62 284 R
62 0 V
873 2060 M
0 218 V
842 2060 M
62 0 V
-62 218 R
62 0 V
353 125 R
0 274 V
-31 -274 R
62 0 V
-62 274 R
62 0 V
22 489 R
0 335 V
-31 -335 R
62 0 V
-62 335 R
62 0 V
30 -767 R
0 291 V
-31 -291 R
62 0 V
-62 291 R
62 0 V
899 2059 M
0 207 V
868 2059 M
62 0 V
stroke 930 2059 M
-62 207 R
62 0 V
330 26 R
0 266 V
-31 -266 R
62 0 V
-62 266 R
62 0 V
22 578 R
0 280 V
-31 -280 R
62 0 V
-62 280 R
62 0 V
30 -694 R
0 289 V
-31 -289 R
62 0 V
-62 289 R
62 0 V
926 2017 M
0 189 V
895 2017 M
62 0 V
-62 189 R
62 0 V
305 421 R
0 236 V
-31 -236 R
62 0 V
-62 236 R
62 0 V
22 258 R
0 310 V
-31 -310 R
62 0 V
-62 310 R
62 0 V
30 -448 R
0 318 V
-31 -318 R
62 0 V
-62 318 R
62 0 V
952 1931 M
0 195 V
921 1931 M
62 0 V
-62 195 R
62 0 V
282 381 R
0 272 V
-31 -272 R
62 0 V
-62 272 R
62 0 V
22 377 R
0 317 V
-31 -317 R
62 0 V
-62 317 R
62 0 V
30 -568 R
0 320 V
-31 -320 R
62 0 V
-62 320 R
62 0 V
979 1966 M
0 189 V
948 1966 M
62 0 V
-62 189 R
62 0 V
258 293 R
0 300 V
-31 -300 R
62 0 V
-62 300 R
62 0 V
22 331 R
0 253 V
-31 -253 R
62 0 V
-62 253 R
62 0 V
30 -478 R
0 282 V
-31 -282 R
62 0 V
-62 282 R
62 0 V
1005 1977 M
0 154 V
974 1977 M
62 0 V
-62 154 R
62 0 V
234 400 R
0 262 V
-31 -262 R
62 0 V
-62 262 R
62 0 V
stroke 1301 2793 M
22 178 R
0 290 V
-31 -290 R
62 0 V
-62 290 R
62 0 V
30 -213 R
0 328 V
-31 -328 R
62 0 V
-62 328 R
62 0 V
1191 1994 M
0 145 V
-31 -145 R
62 0 V
-62 145 R
62 0 V
51 596 R
0 328 V
-31 -328 R
62 0 V
-62 328 R
62 0 V
22 -525 R
0 299 V
-31 -299 R
62 0 V
-62 299 R
62 0 V
30 113 R
0 307 V
-31 -307 R
62 0 V
-62 307 R
62 0 V
1199 1881 M
0 206 V
-31 -206 R
62 0 V
-62 206 R
62 0 V
46 1219 R
0 337 V
-31 -337 R
62 0 V
-62 337 R
62 0 V
22 -847 R
0 320 V
-31 -320 R
62 0 V
-62 320 R
62 0 V
30 -222 R
0 290 V
-31 -290 R
62 0 V
-62 290 R
62 0 V
1204 2124 M
0 192 V
-31 -192 R
62 0 V
-62 192 R
62 0 V
43 1078 R
0 353 V
-31 -353 R
62 0 V
-62 353 R
62 0 V
22 -1180 R
0 254 V
-31 -254 R
62 0 V
-62 254 R
62 0 V
30 11 R
0 299 V
-31 -299 R
62 0 V
-62 299 R
62 0 V
1207 2303 M
0 218 V
-31 -218 R
62 0 V
-62 218 R
62 0 V
40 854 R
0 391 V
-31 -391 R
62 0 V
-62 391 R
62 0 V
25 -1176 R
0 246 V
-31 -246 R
62 0 V
-62 246 R
62 0 V
33 -89 R
0 316 V
stroke 1398 3063 M
-31 -316 R
62 0 V
-62 316 R
62 0 V
1208 2492 M
0 507 V
-31 -507 R
62 0 V
-62 507 R
62 0 V
40 227 R
0 356 V
-31 -356 R
62 0 V
-62 356 R
62 0 V
25 -1156 R
0 228 V
-31 -228 R
62 0 V
-62 228 R
62 0 V
37 88 R
0 262 V
-31 -262 R
62 0 V
-62 262 R
62 0 V
1209 2807 M
0 529 V
-31 -529 R
62 0 V
-62 529 R
62 0 V
41 -44 R
0 326 V
-31 -326 R
62 0 V
-62 326 R
62 0 V
25 -1357 R
0 226 V
-31 -226 R
62 0 V
-62 226 R
62 0 V
40 105 R
0 255 V
-31 -255 R
62 0 V
-62 255 R
62 0 V
-230 -87 R
0 302 V
-31 -302 R
62 0 V
-62 302 R
62 0 V
44 107 R
0 355 V
-31 -355 R
62 0 V
-62 355 R
62 0 V
23 -1155 R
0 228 V
-31 -228 R
62 0 V
-62 228 R
62 0 V
45 -36 R
0 292 V
-31 -292 R
62 0 V
-62 292 R
62 0 V
1210 2424 M
0 425 V
-31 -425 R
62 0 V
-62 425 R
62 0 V
45 594 R
0 344 V
-31 -344 R
62 0 V
-62 344 R
62 0 V
22 -1361 R
0 228 V
-31 -228 R
62 0 V
-62 228 R
62 0 V
49 170 R
0 264 V
-31 -264 R
62 0 V
-62 264 R
62 0 V
1211 2520 M
0 412 V
-31 -412 R
62 0 V
stroke 1242 2520 M
-62 412 R
62 0 V
47 180 R
0 353 V
-31 -353 R
62 0 V
-62 353 R
62 0 V
21 -1002 R
0 244 V
-31 -244 R
62 0 V
-62 244 R
62 0 V
57 -73 R
0 248 V
-31 -248 R
62 0 V
-62 248 R
62 0 V
1212 2630 M
0 217 V
-31 -217 R
62 0 V
-62 217 R
62 0 V
49 329 R
0 354 V
-31 -354 R
62 0 V
-62 354 R
62 0 V
19 -932 R
0 256 V
-31 -256 R
62 0 V
-62 256 R
62 0 V
109 -23 R
0 302 V
-31 -302 R
62 0 V
-62 302 R
62 0 V
1215 2221 M
0 191 V
-31 -191 R
62 0 V
-62 191 R
62 0 V
48 667 R
0 317 V
-31 -317 R
62 0 V
-62 317 R
62 0 V
20 -908 R
0 246 V
-31 -246 R
62 0 V
-62 246 R
62 0 V
1220 2063 M
0 200 V
-31 -200 R
62 0 V
-62 200 R
62 0 V
46 763 R
0 308 V
-31 -308 R
62 0 V
-62 308 R
62 0 V
22 -813 R
0 218 V
-31 -218 R
62 0 V
-62 218 R
62 0 V
1231 2111 M
0 218 V
-31 -218 R
62 0 V
-62 218 R
62 0 V
38 660 R
0 318 V
-31 -318 R
62 0 V
-62 318 R
62 0 V
24 -891 R
0 236 V
-31 -236 R
62 0 V
-62 236 R
62 0 V
740 2035 Dia
1241 2329 Dia
1302 3270 Dia
1361 2732 Dia
793 2163 Dia
1252 2675 Dia
1305 3193 Dia
1363 2873 Dia
846 2163 Dia
1254 2521 Dia
1307 3091 Dia
1366 2860 Dia
873 2169 Dia
1257 2540 Dia
1310 3334 Dia
1371 2879 Dia
899 2163 Dia
1260 2425 Dia
1313 3276 Dia
1374 2867 Dia
926 2112 Dia
1262 2745 Dia
1315 3276 Dia
1376 3142 Dia
952 2028 Dia
1265 2643 Dia
1318 3315 Dia
1379 3065 Dia
979 2060 Dia
1268 2598 Dia
1321 3206 Dia
1382 2995 Dia
1005 2054 Dia
1270 2662 Dia
1323 3116 Dia
1384 3212 Dia
1191 2067 Dia
1273 2899 Dia
1326 2688 Dia
1387 3103 Dia
1199 1984 Dia
1276 3475 Dia
1329 2956 Dia
1390 3039 Dia
1204 2220 Dia
1278 3570 Dia
1331 2694 Dia
1392 2982 Dia
1207 2412 Dia
1278 3570 Dia
1334 2713 Dia
1398 2905 Dia
1208 2745 Dia
1279 3404 Dia
1335 2540 Dia
1403 2873 Dia
1209 3071 Dia
1281 3455 Dia
1337 2374 Dia
1408 2719 Dia
1209 2911 Dia
1284 3347 Dia
1338 2483 Dia
1414 2707 Dia
1210 2636 Dia
1286 3615 Dia
1339 2540 Dia
1419 2956 Dia
1211 2726 Dia
1289 3289 Dia
1341 2585 Dia
1429 2758 Dia
1212 2739 Dia
1292 3353 Dia
1342 2726 Dia
1482 2982 Dia
1215 2316 Dia
1294 3238 Dia
1345 2611 Dia
1220 2163 Dia
1297 3180 Dia
1350 2630 Dia
1231 2220 Dia
1300 3148 Dia
1355 2534 Dia
6327 3636 Dia
1.000 UP
1.000 UL
LT6
1.00 0.65 0.00 C LCb setrgbcolor
LT6
1.00 0.65 0.00 C 6056 3436 M
543 0 V
-543 31 R
0 -62 V
543 62 R
0 -62 V
1178 1979 M
0 120 V
-31 -120 R
62 0 V
-62 120 R
62 0 V
1 810 R
0 301 V
-31 -301 R
62 0 V
-62 301 R
62 0 V
1180 1951 M
0 174 V
-31 -174 R
62 0 V
-62 174 R
62 0 V
-1 662 R
0 271 V
-31 -271 R
62 0 V
-62 271 R
62 0 V
-52 -717 R
0 185 V
-31 -185 R
62 0 V
-62 185 R
62 0 V
-10 147 R
0 256 V
-31 -256 R
62 0 V
-62 256 R
62 0 V
-50 -932 R
0 152 V
-31 -152 R
62 0 V
-62 152 R
62 0 V
-11 684 R
0 278 V
-31 -278 R
62 0 V
-62 278 R
62 0 V
1193 2014 M
0 157 V
-31 -157 R
62 0 V
-62 157 R
62 0 V
-13 576 R
0 267 V
-31 -267 R
62 0 V
-62 267 R
62 0 V
1195 1984 M
0 158 V
-31 -158 R
62 0 V
-62 158 R
62 0 V
-15 649 R
0 272 V
-31 -272 R
62 0 V
-62 272 R
62 0 V
1197 1947 M
0 155 V
-31 -155 R
62 0 V
-62 155 R
62 0 V
-16 560 R
0 259 V
-31 -259 R
62 0 V
-62 259 R
62 0 V
-44 -971 R
0 157 V
-31 -157 R
62 0 V
-62 157 R
62 0 V
-18 620 R
0 279 V
-31 -279 R
62 0 V
-62 279 R
62 0 V
1201 1963 M
0 169 V
-31 -169 R
62 0 V
stroke 1232 1963 M
-62 169 R
62 0 V
-20 569 R
0 268 V
-31 -268 R
62 0 V
-62 268 R
62 0 V
-39 -907 R
0 201 V
-31 -201 R
62 0 V
-62 201 R
62 0 V
-23 405 R
0 276 V
-31 -276 R
62 0 V
-62 276 R
62 0 V
-39 -784 R
0 191 V
-31 -191 R
62 0 V
-62 191 R
62 0 V
-23 293 R
0 266 V
-31 -266 R
62 0 V
-62 266 R
62 0 V
-39 -679 R
0 201 V
-31 -201 R
62 0 V
-62 201 R
62 0 V
-23 287 R
0 280 V
-31 -280 R
62 0 V
-62 280 R
62 0 V
-38 -783 R
0 204 V
-31 -204 R
62 0 V
-62 204 R
62 0 V
-24 260 R
0 270 V
-31 -270 R
62 0 V
-62 270 R
62 0 V
-38 -711 R
0 205 V
-31 -205 R
62 0 V
-62 205 R
62 0 V
-24 311 R
0 272 V
-31 -272 R
62 0 V
-62 272 R
62 0 V
-38 -739 R
0 209 V
-31 -209 R
62 0 V
-62 209 R
62 0 V
-24 -99 R
0 255 V
-31 -255 R
62 0 V
-62 255 R
62 0 V
-37 -455 R
0 199 V
-31 -199 R
62 0 V
-62 199 R
62 0 V
-25 127 R
0 254 V
-31 -254 R
62 0 V
-62 254 R
62 0 V
-37 -367 R
0 230 V
-31 -230 R
62 0 V
-62 230 R
62 0 V
-24 -46 R
0 258 V
-31 -258 R
62 0 V
-62 258 R
62 0 V
stroke 1244 2853 M
-38 -391 R
0 235 V
-31 -235 R
62 0 V
-62 235 R
62 0 V
-24 -55 R
0 256 V
-31 -256 R
62 0 V
-62 256 R
62 0 V
-37 -403 R
0 234 V
-31 -234 R
62 0 V
-62 234 R
62 0 V
-25 -192 R
0 246 V
-31 -246 R
62 0 V
-62 246 R
62 0 V
-37 -246 R
0 228 V
-31 -228 R
62 0 V
-62 228 R
62 0 V
-25 -394 R
0 226 V
-31 -226 R
62 0 V
-62 226 R
62 0 V
-37 -156 R
0 235 V
-31 -235 R
62 0 V
-62 235 R
62 0 V
-24 -290 R
0 226 V
-31 -226 R
62 0 V
-62 226 R
62 0 V
-38 -40 R
0 248 V
-31 -248 R
62 0 V
-62 248 R
62 0 V
-24 -403 R
0 234 V
-31 -234 R
62 0 V
-62 234 R
62 0 V
-37 2 R
0 261 V
-31 -261 R
62 0 V
-62 261 R
62 0 V
-25 -729 R
0 204 V
-31 -204 R
62 0 V
-62 204 R
62 0 V
-37 410 R
0 277 V
-31 -277 R
62 0 V
-62 277 R
62 0 V
-24 -888 R
0 205 V
-31 -205 R
62 0 V
-62 205 R
62 0 V
-38 283 R
0 261 V
-31 -261 R
62 0 V
-62 261 R
62 0 V
-24 -585 R
0 225 V
-31 -225 R
62 0 V
-62 225 R
62 0 V
-37 327 R
0 289 V
-31 -289 R
62 0 V
-62 289 R
62 0 V
-24 -899 R
0 216 V
stroke 1216 2510 M
-31 -216 R
62 0 V
-62 216 R
62 0 V
-38 207 R
0 268 V
-31 -268 R
62 0 V
-62 268 R
62 0 V
1217 1967 M
0 181 V
-31 -181 R
62 0 V
-62 181 R
62 0 V
-39 666 R
0 278 V
-31 -278 R
62 0 V
-62 278 R
62 0 V
1220 1949 M
0 162 V
-31 -162 R
62 0 V
-62 162 R
62 0 V
-41 685 R
0 274 V
-31 -274 R
62 0 V
-62 274 R
62 0 V
1223 2063 M
0 185 V
-31 -185 R
62 0 V
-62 185 R
62 0 V
-44 517 R
0 284 V
-31 -284 R
62 0 V
-62 284 R
62 0 V
-16 -990 R
0 176 V
-31 -176 R
62 0 V
-62 176 R
62 0 V
-46 585 R
0 295 V
-31 -295 R
62 0 V
-62 295 R
62 0 V
1228 2021 M
0 178 V
-31 -178 R
62 0 V
-62 178 R
62 0 V
-49 814 R
0 300 V
-31 -300 R
62 0 V
-62 300 R
62 0 V
1231 2063 M
0 191 V
-31 -191 R
62 0 V
-62 191 R
62 0 V
-52 572 R
0 286 V
-31 -286 R
62 0 V
-62 286 R
62 0 V
-8 -1045 R
0 181 V
-31 -181 R
62 0 V
-62 181 R
62 0 V
-54 585 R
0 293 V
-31 -293 R
62 0 V
-62 293 R
62 0 V
-5 -1104 R
0 191 V
-31 -191 R
62 0 V
-62 191 R
62 0 V
1178 2039 Dia
1210 3059 Dia
1180 2038 Dia
1210 2922 Dia
1189 2433 Dia
1210 2801 Dia
1191 2073 Dia
1211 2972 Dia
1193 2092 Dia
1211 2881 Dia
1195 2063 Dia
1211 2927 Dia
1197 2025 Dia
1212 2792 Dia
1199 2028 Dia
1212 2867 Dia
1201 2048 Dia
1212 2835 Dia
1204 2163 Dia
1212 2806 Dia
1204 2256 Dia
1212 2777 Dia
1204 2332 Dia
1212 2859 Dia
1205 2318 Dia
1212 2815 Dia
1205 2341 Dia
1212 2891 Dia
1205 2393 Dia
1212 2526 Dia
1206 2298 Dia
1212 2651 Dia
1206 2526 Dia
1213 2724 Dia
1206 2579 Dia
1213 2770 Dia
1207 2612 Dia
1213 2660 Dia
1207 2651 Dia
1213 2484 Dia
1207 2558 Dia
1214 2499 Dia
1207 2696 Dia
1214 2534 Dia
1208 2783 Dia
1214 2287 Dia
1208 2937 Dia
1215 2290 Dia
1208 2807 Dia
1215 2465 Dia
1209 3048 Dia
1216 2402 Dia
1209 2851 Dia
1217 2057 Dia
1209 2953 Dia
1220 2030 Dia
1210 2933 Dia
1223 2156 Dia
1210 2907 Dia
1225 2147 Dia
1210 2968 Dia
1228 2110 Dia
1210 3163 Dia
1231 2158 Dia
1210 2969 Dia
1233 2158 Dia
1210 2979 Dia
1236 2117 Dia
6327 3436 Dia
1.000 UL
LTb
740 4799 N
740 640 L
6099 0 V
0 4159 V
-6099 0 V
Z stroke
1.000 UP
1.000 UL
LTb
1.000 UL
LTb
3779 1739 M
63 0 V
2637 0 R
-63 0 V
3779 2783 M
63 0 V
2637 0 R
-63 0 V
4229 904 M
0 63 V
0 1816 R
0 -63 V
5129 904 M
0 63 V
0 1816 R
0 -63 V
6029 904 M
0 63 V
0 1816 R
0 -63 V
stroke
3779 2783 N
0 -1879 V
2700 0 V
0 1879 V
-2700 0 V
Z stroke
LCb setrgbcolor
LTb
LCb setrgbcolor
LTb
1.000 UP
1.000 UL
LTb
1.000 UP
1.000 UL
LT0
0.00 0.00 1.00 C 3824 1791 M
0 606 V
-31 -606 R
62 0 V
-62 606 R
62 0 V
14 -1389 R
0 564 V
-31 -564 R
62 0 V
-62 564 R
62 0 V
14 -449 R
0 564 V
-31 -564 R
62 0 V
-62 564 R
62 0 V
14 -21 R
0 647 V
-31 -647 R
62 0 V
-62 647 R
62 0 V
14 -1273 R
0 501 V
-31 -501 R
62 0 V
-62 501 R
62 0 V
14 -52 R
0 668 V
-31 -668 R
62 0 V
-62 668 R
62 0 V
14 -491 R
0 334 V
-31 -334 R
62 0 V
-62 334 R
62 0 V
-8 -344 R
0 417 V
-31 -417 R
62 0 V
-62 417 R
62 0 V
-9 -365 R
0 334 V
-31 -334 R
62 0 V
-62 334 R
62 0 V
-9 -366 R
0 397 V
-31 -397 R
62 0 V
-62 397 R
62 0 V
-8 -386 R
0 292 V
-31 -292 R
62 0 V
-62 292 R
62 0 V
-8 -229 R
0 334 V
-31 -334 R
62 0 V
-62 334 R
62 0 V
-9 -261 R
0 188 V
-31 -188 R
62 0 V
-62 188 R
62 0 V
-9 -293 R
0 376 V
-31 -376 R
62 0 V
-62 376 R
62 0 V
-8 -470 R
0 313 V
-31 -313 R
62 0 V
-62 313 R
62 0 V
-8 -10 R
0 334 V
-31 -334 R
62 0 V
-62 334 R
62 0 V
-9 -553 R
0 167 V
-31 -167 R
62 0 V
-62 167 R
62 0 V
-8 -324 R
0 292 V
-31 -292 R
62 0 V
stroke 4373 1551 M
-62 292 R
62 0 V
-9 32 R
0 292 V
-31 -292 R
62 0 V
-62 292 R
62 0 V
-9 -115 R
0 355 V
-31 -355 R
62 0 V
-62 355 R
62 0 V
-8 -480 R
0 271 V
-31 -271 R
62 0 V
-62 271 R
62 0 V
-8 -365 R
0 292 V
-31 -292 R
62 0 V
-62 292 R
62 0 V
-9 -125 R
0 313 V
-31 -313 R
62 0 V
-62 313 R
62 0 V
-9 -52 R
0 334 V
-31 -334 R
62 0 V
-62 334 R
62 0 V
-8 -511 R
0 355 V
-31 -355 R
62 0 V
-62 355 R
62 0 V
-8 -606 R
0 334 V
-31 -334 R
62 0 V
-62 334 R
62 0 V
-9 73 R
0 313 V
-31 -313 R
62 0 V
-62 313 R
62 0 V
-8 -574 R
0 313 V
-31 -313 R
62 0 V
-62 313 R
62 0 V
-9 -261 R
0 334 V
-31 -334 R
62 0 V
-62 334 R
62 0 V
-9 -240 R
0 272 V
-31 -272 R
62 0 V
-62 272 R
62 0 V
-8 -386 R
0 313 V
-31 -313 R
62 0 V
-62 313 R
62 0 V
-8 -522 R
0 355 V
-31 -355 R
62 0 V
-62 355 R
62 0 V
-9 -439 R
0 251 V
-31 -251 R
62 0 V
-62 251 R
62 0 V
-9 0 R
0 188 V
-31 -188 R
62 0 V
-62 188 R
62 0 V
-8 -230 R
0 292 V
-31 -292 R
62 0 V
-62 292 R
62 0 V
stroke 4755 2219 M
-8 -292 R
0 230 V
-31 -230 R
62 0 V
-62 230 R
62 0 V
-9 -251 R
0 230 V
-31 -230 R
62 0 V
-62 230 R
62 0 V
-8 -105 R
0 230 V
-31 -230 R
62 0 V
-62 230 R
62 0 V
-9 -334 R
0 230 V
-31 -230 R
62 0 V
-62 230 R
62 0 V
-9 -167 R
0 188 V
-31 -188 R
62 0 V
-62 188 R
62 0 V
-8 -42 R
0 167 V
-31 -167 R
62 0 V
-62 167 R
62 0 V
-8 -178 R
0 293 V
-31 -293 R
62 0 V
-62 293 R
62 0 V
-9 -1107 R
0 1232 V
4873 1311 M
62 0 V
-62 1232 R
62 0 V
104 -867 R
0 1107 V
5008 1676 M
62 0 V
-62 1107 R
62 0 V
3824 2094 BoxF
3869 1290 BoxF
3914 1405 BoxF
3959 1990 BoxF
4004 1290 BoxF
4049 1823 BoxF
4094 1833 BoxF
4117 1864 BoxF
4139 1875 BoxF
4161 1875 BoxF
4184 1833 BoxF
4207 1917 BoxF
4229 1917 BoxF
4251 1906 BoxF
4274 1781 BoxF
4297 2094 BoxF
4319 1791 BoxF
4342 1697 BoxF
4364 2021 BoxF
4386 2230 BoxF
4409 2063 BoxF
4432 1979 BoxF
4454 2157 BoxF
4476 2428 BoxF
4499 2261 BoxF
4522 2000 BoxF
4544 2397 BoxF
4567 2136 BoxF
4589 2198 BoxF
4611 2261 BoxF
4634 2167 BoxF
4657 1979 BoxF
4679 1843 BoxF
4701 2063 BoxF
4724 2073 BoxF
4747 2042 BoxF
4769 2021 BoxF
4792 2146 BoxF
4814 2042 BoxF
4836 2084 BoxF
4859 2219 BoxF
4882 2271 BoxF
4904 1927 BoxF
5039 2230 BoxF
1.000 UP
1.000 UL
LT1
0.00 0.50 1.00 C 3779 957 M
0 354 V
3748 957 M
62 0 V
-62 354 R
62 0 V
82 -140 R
0 238 V
-31 -238 R
62 0 V
-62 238 R
62 0 V
81 -408 R
0 244 V
-31 -244 R
62 0 V
-62 244 R
62 0 V
82 -207 R
0 233 V
-31 -233 R
62 0 V
-62 233 R
62 0 V
81 -188 R
0 268 V
-31 -268 R
62 0 V
-62 268 R
62 0 V
82 -441 R
0 218 V
4311 910 M
62 0 V
-62 218 R
62 0 V
81 -156 R
0 218 V
4423 972 M
62 0 V
-62 218 R
62 0 V
82 -169 R
0 226 V
-31 -226 R
62 0 V
-62 226 R
62 0 V
81 -299 R
0 225 V
4648 948 M
62 0 V
-62 225 R
62 0 V
82 -19 R
0 272 V
-31 -272 R
62 0 V
-62 272 R
62 0 V
36 -512 R
0 293 V
4828 914 M
62 0 V
-62 293 R
62 0 V
3779 1134 TriU
3892 1290 TriU
4004 1123 TriU
4117 1155 TriU
4229 1217 TriU
4342 1019 TriU
4454 1081 TriU
4567 1134 TriU
4679 1061 TriU
4792 1290 TriU
4859 1061 TriU
1.000 UP
1.000 UL
LT2
0.00 0.75 0.00 C 4814 1108 M
0 365 V
-31 -365 R
62 0 V
-62 365 R
62 0 V
14 -66 R
0 372 V
-31 -372 R
62 0 V
-62 372 R
62 0 V
14 -344 R
0 378 V
-31 -378 R
62 0 V
-62 378 R
62 0 V
14 -332 R
0 391 V
-31 -391 R
62 0 V
-62 391 R
62 0 V
14 -391 R
0 391 V
-31 -391 R
62 0 V
-62 391 R
62 0 V
14 -758 R
0 394 V
-31 -394 R
62 0 V
-62 394 R
62 0 V
14 -324 R
0 401 V
-31 -401 R
62 0 V
-62 401 R
62 0 V
14 -394 R
0 407 V
-31 -407 R
62 0 V
-62 407 R
62 0 V
14 -306 R
0 330 V
-31 -330 R
62 0 V
-62 330 R
62 0 V
14 -427 R
0 336 V
-31 -336 R
62 0 V
-62 336 R
62 0 V
14 -588 R
0 340 V
5233 943 M
62 0 V
-62 340 R
62 0 V
14 -102 R
0 344 V
-31 -344 R
62 0 V
-62 344 R
62 0 V
14 -441 R
0 350 V
-31 -350 R
62 0 V
-62 350 R
62 0 V
14 -227 R
0 355 V
-31 -355 R
62 0 V
-62 355 R
62 0 V
14 -490 R
0 165 V
-31 -165 R
62 0 V
-62 165 R
62 0 V
14 -72 R
0 167 V
-31 -167 R
62 0 V
-62 167 R
62 0 V
14 20 R
0 169 V
-31 -169 R
62 0 V
-62 169 R
62 0 V
14 -240 R
0 227 V
-31 -227 R
62 0 V
stroke 5610 1281 M
-62 227 R
62 0 V
14 -270 R
0 230 V
-31 -230 R
62 0 V
-62 230 R
62 0 V
14 -295 R
0 234 V
-31 -234 R
62 0 V
-62 234 R
62 0 V
14 -203 R
0 236 V
-31 -236 R
62 0 V
-62 236 R
62 0 V
14 -399 R
0 165 V
-31 -165 R
62 0 V
-62 165 R
62 0 V
14 63 R
0 167 V
-31 -167 R
62 0 V
-62 167 R
62 0 V
14 -263 R
0 192 V
-31 -192 R
62 0 V
-62 192 R
62 0 V
14 -224 R
0 194 V
-31 -194 R
62 0 V
-62 194 R
62 0 V
14 -132 R
0 196 V
-31 -196 R
62 0 V
-62 196 R
62 0 V
14 -291 R
0 198 V
-31 -198 R
62 0 V
-62 198 R
62 0 V
14 -37 R
0 272 V
-31 -272 R
62 0 V
-62 272 R
62 0 V
14 -398 R
0 274 V
-31 -274 R
62 0 V
-62 274 R
62 0 V
14 -349 R
0 278 V
-31 -278 R
62 0 V
-62 278 R
62 0 V
14 -399 R
0 499 V
6133 947 M
62 0 V
-62 499 R
62 0 V
4814 1290 CircleF
4859 1593 CircleF
4904 1624 CircleF
4949 1676 CircleF
4994 1676 CircleF
5039 1311 CircleF
5084 1384 CircleF
5129 1395 CircleF
5174 1457 CircleF
5219 1363 CircleF
5264 1113 CircleF
5309 1353 CircleF
5354 1259 CircleF
5399 1384 CircleF
5444 1155 CircleF
5489 1248 CircleF
5534 1436 CircleF
5579 1395 CircleF
5624 1353 CircleF
5669 1290 CircleF
5714 1322 CircleF
5759 1123 CircleF
5804 1353 CircleF
5849 1269 CircleF
5894 1238 CircleF
5939 1301 CircleF
5984 1207 CircleF
6029 1405 CircleF
6074 1280 CircleF
6119 1207 CircleF
6164 1196 CircleF
1.000 UP
1.000 UL
LT3
0.55 0.00 0.00 C 6271 1225 M
0 174 V
-31 -174 R
62 0 V
-62 174 R
62 0 V
6178 1124 M
0 167 V
-31 -167 R
62 0 V
-62 167 R
62 0 V
6027 1133 M
0 162 V
-31 -162 R
62 0 V
-62 162 R
62 0 V
5773 1148 M
0 159 V
-31 -159 R
62 0 V
-62 159 R
62 0 V
-504 -81 R
0 141 V
-31 -141 R
62 0 V
-62 141 R
62 0 V
4850 1202 M
0 134 V
-31 -134 R
62 0 V
-62 134 R
62 0 V
-218 -89 R
0 143 V
-31 -143 R
62 0 V
-62 143 R
62 0 V
6271 1312 TriD
6178 1208 TriD
6027 1214 TriD
5773 1228 TriD
5300 1297 TriD
4850 1269 TriD
4663 1318 TriD
1.000 UP
1.000 UL
LT4
0.65 0.16 0.16 C 1.000 UP
1.000 UL
LT5
1.00 0.65 0.00 C 1.000 UL
LTb
3779 2783 N
0 -1879 V
2700 0 V
0 1879 V
-2700 0 V
Z stroke
1.000 UP
1.000 UL
LTb
stroke
grestore
end
showpage
  }}%
  \put(6029,804){\makebox(0,0){\strut{} 10}}%
  \put(5129,804){\makebox(0,0){\strut{} 8}}%
  \put(4229,804){\makebox(0,0){\strut{} 6}}%
  \put(3719,2783){\makebox(0,0)[r]{\strut{} 5}}%
  \put(3719,1739){\makebox(0,0)[r]{\strut{} 4}}%
  \put(5936,3436){\makebox(0,0)[r]{\strut{}BES 2006}}%
  \put(5936,3636){\makebox(0,0)[r]{\strut{}BES 2002}}%
  \put(5936,3836){\makebox(0,0)[r]{\strut{}CLEO 2007}}%
  \put(5936,4036){\makebox(0,0)[r]{\strut{}MD-1 1996}}%
  \put(5936,4236){\makebox(0,0)[r]{\strut{}ChrystalBall 1990}}%
  \put(5936,4436){\makebox(0,0)[r]{\strut{}PLUTO}}%
  \put(5936,4636){\makebox(0,0)[r]{\strut{}MARKI 1986}}%
  \put(3789,140){\makebox(0,0){\strut{}$\sqrt{s}$ (in GeV)}}%
  \put(160,2719){\makebox(0,0){\strut{}$R_{exp.}$}}%
  \put(6839,440){\makebox(0,0){\strut{} 25}}%
  \put(5513,440){\makebox(0,0){\strut{} 20}}%
  \put(4187,440){\makebox(0,0){\strut{} 15}}%
  \put(2861,440){\makebox(0,0){\strut{} 10}}%
  \put(1536,440){\makebox(0,0){\strut{} 5}}%
  \put(620,4479){\makebox(0,0)[r]{\strut{} 6}}%
  \put(620,3839){\makebox(0,0)[r]{\strut{} 5}}%
  \put(620,3199){\makebox(0,0)[r]{\strut{} 4}}%
  \put(620,2560){\makebox(0,0)[r]{\strut{} 3}}%
  \put(620,1920){\makebox(0,0)[r]{\strut{} 2}}%
  \put(620,1280){\makebox(0,0)[r]{\strut{} 1}}%
  \put(620,640){\makebox(0,0)[r]{\strut{} 0}}%
\end{picture}%
\endgroup
 

%% file: ExtrapolationSTotal.tex
\begingroup%
\makeatletter%
\newcommand{\GNUPLOTspecial}{%
  \@sanitize\catcode`\%=14\relax\special}%
\setlength{\unitlength}{0.0500bp}%
\begin{picture}(7200,5040)(0,0)%
  {\GNUPLOTspecial{"
/gnudict 256 dict def
gnudict begin
%
%
/Color true def
/Blacktext true def
/Solid true def
/Dashlength 1 def
/Landscape false def
/Level1 false def
/Rounded false def
/ClipToBoundingBox false def
/TransparentPatterns false def
/gnulinewidth 5.000 def
/userlinewidth gnulinewidth def
/Gamma 1.0 def
/vshift -66 def
/dl1 {
  10.0 Dashlength mul mul
  Rounded { currentlinewidth 0.75 mul sub dup 0 le { pop 0.01 } if } if
} def
/dl2 {
  10.0 Dashlength mul mul
  Rounded { currentlinewidth 0.75 mul add } if
} def
/hpt_ 31.5 def
/vpt_ 31.5 def
/hpt hpt_ def
/vpt vpt_ def
Level1 {} {
/SDict 10 dict def
systemdict /pdfmark known not {
  userdict /pdfmark systemdict /cleartomark get put
} if
SDict begin [
  /Title (ExtrapolationSTotal.tex)
  /Subject (gnuplot plot)
  /Creator (gnuplot 4.4 patchlevel 3)
  /Author (dg)
  /CreationDate (Thu Jan 31 15:54:31 2013)
  /DOCINFO pdfmark
end
} ifelse
/doclip {
  ClipToBoundingBox {
    newpath 0 0 moveto 360 0 lineto 360 252 lineto 0 252 lineto closepath
    clip
  } if
} def
%
%
%
/M {moveto} bind def
/L {lineto} bind def
/R {rmoveto} bind def
/V {rlineto} bind def
/N {newpath moveto} bind def
/Z {closepath} bind def
/C {setrgbcolor} bind def
/f {rlineto fill} bind def
/g {setgray} bind def
/Gshow {show} def   
/vpt2 vpt 2 mul def
/hpt2 hpt 2 mul def
/Lshow {currentpoint stroke M 0 vshift R 
	Blacktext {gsave 0 setgray show grestore} {show} ifelse} def
/Rshow {currentpoint stroke M dup stringwidth pop neg vshift R
	Blacktext {gsave 0 setgray show grestore} {show} ifelse} def
/Cshow {currentpoint stroke M dup stringwidth pop -2 div vshift R 
	Blacktext {gsave 0 setgray show grestore} {show} ifelse} def
/UP {dup vpt_ mul /vpt exch def hpt_ mul /hpt exch def
  /hpt2 hpt 2 mul def /vpt2 vpt 2 mul def} def
/DL {Color {setrgbcolor Solid {pop []} if 0 setdash}
 {pop pop pop 0 setgray Solid {pop []} if 0 setdash} ifelse} def
/BL {stroke userlinewidth 2 mul setlinewidth
	Rounded {1 setlinejoin 1 setlinecap} if} def
/AL {stroke userlinewidth 2 div setlinewidth
	Rounded {1 setlinejoin 1 setlinecap} if} def
/UL {dup gnulinewidth mul /userlinewidth exch def
	dup 1 lt {pop 1} if 10 mul /udl exch def} def
/PL {stroke userlinewidth setlinewidth
	Rounded {1 setlinejoin 1 setlinecap} if} def
3.8 setmiterlimit
/LCw {1 1 1} def
/LCb {0 0 0} def
/LCa {0 0 0} def
/LC0 {1 0 0} def
/LC1 {0 1 0} def
/LC2 {0 0 1} def
/LC3 {1 0 1} def
/LC4 {0 1 1} def
/LC5 {1 1 0} def
/LC6 {0 0 0} def
/LC7 {1 0.3 0} def
/LC8 {0.5 0.5 0.5} def
/LTw {PL [] 1 setgray} def
/LTb {BL [] LCb DL} def
/LTa {AL [1 udl mul 2 udl mul] 0 setdash LCa setrgbcolor} def
/LT0 {PL [] LC0 DL} def
/LT1 {PL [4 dl1 2 dl2] LC1 DL} def
/LT2 {PL [2 dl1 3 dl2] LC2 DL} def
/LT3 {PL [1 dl1 1.5 dl2] LC3 DL} def
/LT4 {PL [6 dl1 2 dl2 1 dl1 2 dl2] LC4 DL} def
/LT5 {PL [3 dl1 3 dl2 1 dl1 3 dl2] LC5 DL} def
/LT6 {PL [2 dl1 2 dl2 2 dl1 6 dl2] LC6 DL} def
/LT7 {PL [1 dl1 2 dl2 6 dl1 2 dl2 1 dl1 2 dl2] LC7 DL} def
/LT8 {PL [2 dl1 2 dl2 2 dl1 2 dl2 2 dl1 2 dl2 2 dl1 4 dl2] LC8 DL} def
/Pnt {stroke [] 0 setdash gsave 1 setlinecap M 0 0 V stroke grestore} def
/Dia {stroke [] 0 setdash 2 copy vpt add M
  hpt neg vpt neg V hpt vpt neg V
  hpt vpt V hpt neg vpt V closepath stroke
  Pnt} def
/Pls {stroke [] 0 setdash vpt sub M 0 vpt2 V
  currentpoint stroke M
  hpt neg vpt neg R hpt2 0 V stroke
 } def
/Box {stroke [] 0 setdash 2 copy exch hpt sub exch vpt add M
  0 vpt2 neg V hpt2 0 V 0 vpt2 V
  hpt2 neg 0 V closepath stroke
  Pnt} def
/Crs {stroke [] 0 setdash exch hpt sub exch vpt add M
  hpt2 vpt2 neg V currentpoint stroke M
  hpt2 neg 0 R hpt2 vpt2 V stroke} def
/TriU {stroke [] 0 setdash 2 copy vpt 1.12 mul add M
  hpt neg vpt -1.62 mul V
  hpt 2 mul 0 V
  hpt neg vpt 1.62 mul V closepath stroke
  Pnt} def
/Star {2 copy Pls Crs} def
/BoxF {stroke [] 0 setdash exch hpt sub exch vpt add M
  0 vpt2 neg V hpt2 0 V 0 vpt2 V
  hpt2 neg 0 V closepath fill} def
/TriUF {stroke [] 0 setdash vpt 1.12 mul add M
  hpt neg vpt -1.62 mul V
  hpt 2 mul 0 V
  hpt neg vpt 1.62 mul V closepath fill} def
/TriD {stroke [] 0 setdash 2 copy vpt 1.12 mul sub M
  hpt neg vpt 1.62 mul V
  hpt 2 mul 0 V
  hpt neg vpt -1.62 mul V closepath stroke
  Pnt} def
/TriDF {stroke [] 0 setdash vpt 1.12 mul sub M
  hpt neg vpt 1.62 mul V
  hpt 2 mul 0 V
  hpt neg vpt -1.62 mul V closepath fill} def
/DiaF {stroke [] 0 setdash vpt add M
  hpt neg vpt neg V hpt vpt neg V
  hpt vpt V hpt neg vpt V closepath fill} def
/Pent {stroke [] 0 setdash 2 copy gsave
  translate 0 hpt M 4 {72 rotate 0 hpt L} repeat
  closepath stroke grestore Pnt} def
/PentF {stroke [] 0 setdash gsave
  translate 0 hpt M 4 {72 rotate 0 hpt L} repeat
  closepath fill grestore} def
/Circle {stroke [] 0 setdash 2 copy
  hpt 0 360 arc stroke Pnt} def
/CircleF {stroke [] 0 setdash hpt 0 360 arc fill} def
/C0 {BL [] 0 setdash 2 copy moveto vpt 90 450 arc} bind def
/C1 {BL [] 0 setdash 2 copy moveto
	2 copy vpt 0 90 arc closepath fill
	vpt 0 360 arc closepath} bind def
/C2 {BL [] 0 setdash 2 copy moveto
	2 copy vpt 90 180 arc closepath fill
	vpt 0 360 arc closepath} bind def
/C3 {BL [] 0 setdash 2 copy moveto
	2 copy vpt 0 180 arc closepath fill
	vpt 0 360 arc closepath} bind def
/C4 {BL [] 0 setdash 2 copy moveto
	2 copy vpt 180 270 arc closepath fill
	vpt 0 360 arc closepath} bind def
/C5 {BL [] 0 setdash 2 copy moveto
	2 copy vpt 0 90 arc
	2 copy moveto
	2 copy vpt 180 270 arc closepath fill
	vpt 0 360 arc} bind def
/C6 {BL [] 0 setdash 2 copy moveto
	2 copy vpt 90 270 arc closepath fill
	vpt 0 360 arc closepath} bind def
/C7 {BL [] 0 setdash 2 copy moveto
	2 copy vpt 0 270 arc closepath fill
	vpt 0 360 arc closepath} bind def
/C8 {BL [] 0 setdash 2 copy moveto
	2 copy vpt 270 360 arc closepath fill
	vpt 0 360 arc closepath} bind def
/C9 {BL [] 0 setdash 2 copy moveto
	2 copy vpt 270 450 arc closepath fill
	vpt 0 360 arc closepath} bind def
/C10 {BL [] 0 setdash 2 copy 2 copy moveto vpt 270 360 arc closepath fill
	2 copy moveto
	2 copy vpt 90 180 arc closepath fill
	vpt 0 360 arc closepath} bind def
/C11 {BL [] 0 setdash 2 copy moveto
	2 copy vpt 0 180 arc closepath fill
	2 copy moveto
	2 copy vpt 270 360 arc closepath fill
	vpt 0 360 arc closepath} bind def
/C12 {BL [] 0 setdash 2 copy moveto
	2 copy vpt 180 360 arc closepath fill
	vpt 0 360 arc closepath} bind def
/C13 {BL [] 0 setdash 2 copy moveto
	2 copy vpt 0 90 arc closepath fill
	2 copy moveto
	2 copy vpt 180 360 arc closepath fill
	vpt 0 360 arc closepath} bind def
/C14 {BL [] 0 setdash 2 copy moveto
	2 copy vpt 90 360 arc closepath fill
	vpt 0 360 arc} bind def
/C15 {BL [] 0 setdash 2 copy vpt 0 360 arc closepath fill
	vpt 0 360 arc closepath} bind def
/Rec {newpath 4 2 roll moveto 1 index 0 rlineto 0 exch rlineto
	neg 0 rlineto closepath} bind def
/Square {dup Rec} bind def
/Bsquare {vpt sub exch vpt sub exch vpt2 Square} bind def
/S0 {BL [] 0 setdash 2 copy moveto 0 vpt rlineto BL Bsquare} bind def
/S1 {BL [] 0 setdash 2 copy vpt Square fill Bsquare} bind def
/S2 {BL [] 0 setdash 2 copy exch vpt sub exch vpt Square fill Bsquare} bind def
/S3 {BL [] 0 setdash 2 copy exch vpt sub exch vpt2 vpt Rec fill Bsquare} bind def
/S4 {BL [] 0 setdash 2 copy exch vpt sub exch vpt sub vpt Square fill Bsquare} bind def
/S5 {BL [] 0 setdash 2 copy 2 copy vpt Square fill
	exch vpt sub exch vpt sub vpt Square fill Bsquare} bind def
/S6 {BL [] 0 setdash 2 copy exch vpt sub exch vpt sub vpt vpt2 Rec fill Bsquare} bind def
/S7 {BL [] 0 setdash 2 copy exch vpt sub exch vpt sub vpt vpt2 Rec fill
	2 copy vpt Square fill Bsquare} bind def
/S8 {BL [] 0 setdash 2 copy vpt sub vpt Square fill Bsquare} bind def
/S9 {BL [] 0 setdash 2 copy vpt sub vpt vpt2 Rec fill Bsquare} bind def
/S10 {BL [] 0 setdash 2 copy vpt sub vpt Square fill 2 copy exch vpt sub exch vpt Square fill
	Bsquare} bind def
/S11 {BL [] 0 setdash 2 copy vpt sub vpt Square fill 2 copy exch vpt sub exch vpt2 vpt Rec fill
	Bsquare} bind def
/S12 {BL [] 0 setdash 2 copy exch vpt sub exch vpt sub vpt2 vpt Rec fill Bsquare} bind def
/S13 {BL [] 0 setdash 2 copy exch vpt sub exch vpt sub vpt2 vpt Rec fill
	2 copy vpt Square fill Bsquare} bind def
/S14 {BL [] 0 setdash 2 copy exch vpt sub exch vpt sub vpt2 vpt Rec fill
	2 copy exch vpt sub exch vpt Square fill Bsquare} bind def
/S15 {BL [] 0 setdash 2 copy Bsquare fill Bsquare} bind def
/D0 {gsave translate 45 rotate 0 0 S0 stroke grestore} bind def
/D1 {gsave translate 45 rotate 0 0 S1 stroke grestore} bind def
/D2 {gsave translate 45 rotate 0 0 S2 stroke grestore} bind def
/D3 {gsave translate 45 rotate 0 0 S3 stroke grestore} bind def
/D4 {gsave translate 45 rotate 0 0 S4 stroke grestore} bind def
/D5 {gsave translate 45 rotate 0 0 S5 stroke grestore} bind def
/D6 {gsave translate 45 rotate 0 0 S6 stroke grestore} bind def
/D7 {gsave translate 45 rotate 0 0 S7 stroke grestore} bind def
/D8 {gsave translate 45 rotate 0 0 S8 stroke grestore} bind def
/D9 {gsave translate 45 rotate 0 0 S9 stroke grestore} bind def
/D10 {gsave translate 45 rotate 0 0 S10 stroke grestore} bind def
/D11 {gsave translate 45 rotate 0 0 S11 stroke grestore} bind def
/D12 {gsave translate 45 rotate 0 0 S12 stroke grestore} bind def
/D13 {gsave translate 45 rotate 0 0 S13 stroke grestore} bind def
/D14 {gsave translate 45 rotate 0 0 S14 stroke grestore} bind def
/D15 {gsave translate 45 rotate 0 0 S15 stroke grestore} bind def
/DiaE {stroke [] 0 setdash vpt add M
  hpt neg vpt neg V hpt vpt neg V
  hpt vpt V hpt neg vpt V closepath stroke} def
/BoxE {stroke [] 0 setdash exch hpt sub exch vpt add M
  0 vpt2 neg V hpt2 0 V 0 vpt2 V
  hpt2 neg 0 V closepath stroke} def
/TriUE {stroke [] 0 setdash vpt 1.12 mul add M
  hpt neg vpt -1.62 mul V
  hpt 2 mul 0 V
  hpt neg vpt 1.62 mul V closepath stroke} def
/TriDE {stroke [] 0 setdash vpt 1.12 mul sub M
  hpt neg vpt 1.62 mul V
  hpt 2 mul 0 V
  hpt neg vpt -1.62 mul V closepath stroke} def
/PentE {stroke [] 0 setdash gsave
  translate 0 hpt M 4 {72 rotate 0 hpt L} repeat
  closepath stroke grestore} def
/CircE {stroke [] 0 setdash 
  hpt 0 360 arc stroke} def
/Opaque {gsave closepath 1 setgray fill grestore 0 setgray closepath} def
/DiaW {stroke [] 0 setdash vpt add M
  hpt neg vpt neg V hpt vpt neg V
  hpt vpt V hpt neg vpt V Opaque stroke} def
/BoxW {stroke [] 0 setdash exch hpt sub exch vpt add M
  0 vpt2 neg V hpt2 0 V 0 vpt2 V
  hpt2 neg 0 V Opaque stroke} def
/TriUW {stroke [] 0 setdash vpt 1.12 mul add M
  hpt neg vpt -1.62 mul V
  hpt 2 mul 0 V
  hpt neg vpt 1.62 mul V Opaque stroke} def
/TriDW {stroke [] 0 setdash vpt 1.12 mul sub M
  hpt neg vpt 1.62 mul V
  hpt 2 mul 0 V
  hpt neg vpt -1.62 mul V Opaque stroke} def
/PentW {stroke [] 0 setdash gsave
  translate 0 hpt M 4 {72 rotate 0 hpt L} repeat
  Opaque stroke grestore} def
/CircW {stroke [] 0 setdash 
  hpt 0 360 arc Opaque stroke} def
/BoxFill {gsave Rec 1 setgray fill grestore} def
/Density {
  /Fillden exch def
  currentrgbcolor
  /ColB exch def /ColG exch def /ColR exch def
  /ColR ColR Fillden mul Fillden sub 1 add def
  /ColG ColG Fillden mul Fillden sub 1 add def
  /ColB ColB Fillden mul Fillden sub 1 add def
  ColR ColG ColB setrgbcolor} def
/BoxColFill {gsave Rec PolyFill} def
/PolyFill {gsave Density fill grestore grestore} def
/h {rlineto rlineto rlineto gsave closepath fill grestore} bind def
%
%
/PatternFill {gsave /PFa [ 9 2 roll ] def
  PFa 0 get PFa 2 get 2 div add PFa 1 get PFa 3 get 2 div add translate
  PFa 2 get -2 div PFa 3 get -2 div PFa 2 get PFa 3 get Rec
  gsave 1 setgray fill grestore clip
  currentlinewidth 0.5 mul setlinewidth
  /PFs PFa 2 get dup mul PFa 3 get dup mul add sqrt def
  0 0 M PFa 5 get rotate PFs -2 div dup translate
  0 1 PFs PFa 4 get div 1 add floor cvi
	{PFa 4 get mul 0 M 0 PFs V} for
  0 PFa 6 get ne {
	0 1 PFs PFa 4 get div 1 add floor cvi
	{PFa 4 get mul 0 2 1 roll M PFs 0 V} for
 } if
  stroke grestore} def
/languagelevel where
 {pop languagelevel} {1} ifelse
 2 lt
	{/InterpretLevel1 true def}
	{/InterpretLevel1 Level1 def}
 ifelse
%
%
/Level2PatternFill {
/Tile8x8 {/PaintType 2 /PatternType 1 /TilingType 1 /BBox [0 0 8 8] /XStep 8 /YStep 8}
	bind def
/KeepColor {currentrgbcolor [/Pattern /DeviceRGB] setcolorspace} bind def
<< Tile8x8
 /PaintProc {0.5 setlinewidth pop 0 0 M 8 8 L 0 8 M 8 0 L stroke} 
>> matrix makepattern
/Pat1 exch def
<< Tile8x8
 /PaintProc {0.5 setlinewidth pop 0 0 M 8 8 L 0 8 M 8 0 L stroke
	0 4 M 4 8 L 8 4 L 4 0 L 0 4 L stroke}
>> matrix makepattern
/Pat2 exch def
<< Tile8x8
 /PaintProc {0.5 setlinewidth pop 0 0 M 0 8 L
	8 8 L 8 0 L 0 0 L fill}
>> matrix makepattern
/Pat3 exch def
<< Tile8x8
 /PaintProc {0.5 setlinewidth pop -4 8 M 8 -4 L
	0 12 M 12 0 L stroke}
>> matrix makepattern
/Pat4 exch def
<< Tile8x8
 /PaintProc {0.5 setlinewidth pop -4 0 M 8 12 L
	0 -4 M 12 8 L stroke}
>> matrix makepattern
/Pat5 exch def
<< Tile8x8
 /PaintProc {0.5 setlinewidth pop -2 8 M 4 -4 L
	0 12 M 8 -4 L 4 12 M 10 0 L stroke}
>> matrix makepattern
/Pat6 exch def
<< Tile8x8
 /PaintProc {0.5 setlinewidth pop -2 0 M 4 12 L
	0 -4 M 8 12 L 4 -4 M 10 8 L stroke}
>> matrix makepattern
/Pat7 exch def
<< Tile8x8
 /PaintProc {0.5 setlinewidth pop 8 -2 M -4 4 L
	12 0 M -4 8 L 12 4 M 0 10 L stroke}
>> matrix makepattern
/Pat8 exch def
<< Tile8x8
 /PaintProc {0.5 setlinewidth pop 0 -2 M 12 4 L
	-4 0 M 12 8 L -4 4 M 8 10 L stroke}
>> matrix makepattern
/Pat9 exch def
/Pattern1 {PatternBgnd KeepColor Pat1 setpattern} bind def
/Pattern2 {PatternBgnd KeepColor Pat2 setpattern} bind def
/Pattern3 {PatternBgnd KeepColor Pat3 setpattern} bind def
/Pattern4 {PatternBgnd KeepColor Landscape {Pat5} {Pat4} ifelse setpattern} bind def
/Pattern5 {PatternBgnd KeepColor Landscape {Pat4} {Pat5} ifelse setpattern} bind def
/Pattern6 {PatternBgnd KeepColor Landscape {Pat9} {Pat6} ifelse setpattern} bind def
/Pattern7 {PatternBgnd KeepColor Landscape {Pat8} {Pat7} ifelse setpattern} bind def
} def
%
%
%
/PatternBgnd {
  TransparentPatterns {} {gsave 1 setgray fill grestore} ifelse
} def
%
%
/Level1PatternFill {
/Pattern1 {0.250 Density} bind def
/Pattern2 {0.500 Density} bind def
/Pattern3 {0.750 Density} bind def
/Pattern4 {0.125 Density} bind def
/Pattern5 {0.375 Density} bind def
/Pattern6 {0.625 Density} bind def
/Pattern7 {0.875 Density} bind def
} def
%
%
Level1 {Level1PatternFill} {Level2PatternFill} ifelse
/Symbol-Oblique /Symbol findfont [1 0 .167 1 0 0] makefont
dup length dict begin {1 index /FID eq {pop pop} {def} ifelse} forall
currentdict end definefont pop
end
gnudict begin
gsave
doclip
0 0 translate
0.050 0.050 scale
0 setgray
newpath
1.000 UL
LTb
980 640 M
63 0 V
5796 0 R
-63 0 V
980 1056 M
63 0 V
5796 0 R
-63 0 V
980 1472 M
63 0 V
5796 0 R
-63 0 V
980 1888 M
63 0 V
5796 0 R
-63 0 V
980 2304 M
63 0 V
5796 0 R
-63 0 V
980 2720 M
63 0 V
5796 0 R
-63 0 V
980 3135 M
63 0 V
5796 0 R
-63 0 V
980 3551 M
63 0 V
5796 0 R
-63 0 V
980 3967 M
63 0 V
5796 0 R
-63 0 V
980 4383 M
63 0 V
5796 0 R
-63 0 V
980 4799 M
63 0 V
5796 0 R
-63 0 V
980 640 M
0 63 V
0 4096 R
0 -63 V
1957 640 M
0 63 V
0 4096 R
0 -63 V
2933 640 M
0 63 V
0 4096 R
0 -63 V
3910 640 M
0 63 V
0 4096 R
0 -63 V
4886 640 M
0 63 V
0 4096 R
0 -63 V
5863 640 M
0 63 V
0 4096 R
0 -63 V
6839 640 M
0 63 V
0 4096 R
0 -63 V
stroke
980 4799 N
980 640 L
5859 0 V
0 4159 V
-5859 0 V
Z stroke
LCb setrgbcolor
LTb
LCb setrgbcolor
LTb
1.000 UP
1.000 UL
LTb
1.000 UL
LT0
1.00 0.00 0.00 C gsave 1175 2490 N 0 -111 V 49 18 V 0 102 V 0.15 PolyFill
gsave 1224 2499 N 0 -102 V 244 64 V 0 67 V 0.15 PolyFill
gsave 1468 2528 N 0 -67 V 244 46 V 0 31 V 0.15 PolyFill
gsave 1712 2538 N 0 -31 V 245 1 V 0 27 V 0.15 PolyFill
gsave 1957 2535 N 0 -27 V 244 -7 V 0 23 V 0.15 PolyFill
gsave 2201 2524 N 0 -23 V 244 -12 V 0 20 V 0.15 PolyFill
gsave 2445 2509 N 0 -20 V 244 -12 V 0 15 V 0.15 PolyFill
gsave 2689 2492 N 0 -15 V 209 -2 V 0 0 V 0.15 PolyFill
gsave 2898 2475 N 0 0 V 35 -17 V 0 15 V 0.15 PolyFill
gsave 2933 2473 N 0 -15 V 224 -3 V 0 0 V 0.15 PolyFill
gsave 3157 2455 N 0 0 V 20 -1 V 0 -1 V 0.15 PolyFill
gsave 3177 2453 N 0 1 V 40 -4 V 0 0 V 0.15 PolyFill
gsave 3217 2450 N 0 0 V 9 -1 V 0 0 V 0.15 PolyFill
gsave 3226 2449 N 0 0 V 5 0 V 0 0 V 0.15 PolyFill
gsave 3231 2449 N 0 0 V 73 -2 V 0 -4 V 0.15 PolyFill
gsave 3304 2443 N 0 4 V 117 -7 V 0 -7 V 0.15 PolyFill
gsave 3421 2433 N 0 7 V 98 -7 V 0 -7 V 0.15 PolyFill
gsave 3519 2426 N 0 7 V 98 -7 V 0 -8 V 0.15 PolyFill
gsave 3617 2418 N 0 8 V 97 -7 V 0 -9 V 0.15 PolyFill
gsave 3714 2410 N 0 9 V 98 -7 V 0 -9 V 0.15 PolyFill
gsave 3812 2403 N 0 9 V 98 -7 V 0 -10 V 0.15 PolyFill
gsave 3910 2395 N 0 10 V 97 -7 V 0 -10 V 0.15 PolyFill
gsave 4007 2388 N 0 10 V 98 -7 V 0 -10 V 0.15 PolyFill
gsave 4105 2381 N 0 10 V 97 -7 V 0 -10 V 0.15 PolyFill
gsave 4202 2374 N 0 10 V 79 -6 V 0 -10 V 0.15 PolyFill
gsave 4281 2368 N 0 10 V 19 -1 V 0 -11 V 0.15 PolyFill
gsave 4300 2366 N 0 11 V 98 -7 V 0 -10 V 0.15 PolyFill
gsave 4398 2360 N 0 10 V 97 -7 V 0 -10 V 0.15 PolyFill
gsave 4495 2353 N 0 10 V 98 -6 V 0 -11 V 0.15 PolyFill
gsave 4593 2346 N 0 11 V 98 -7 V 0 -10 V 0.15 PolyFill
gsave 4691 2340 N 0 10 V 97 -6 V 0 -11 V 0.15 PolyFill
gsave 4788 2333 N 0 11 V 98 -6 V 0 -11 V 0.15 PolyFill
gsave 4886 2327 N 0 11 V 98 -7 V 0 -10 V 0.15 PolyFill
gsave 4984 2321 N 0 10 V 97 -6 V 0 -10 V 0.15 PolyFill
gsave 5081 2315 N 0 10 V 98 -6 V 0 -10 V 0.15 PolyFill
gsave 5179 2309 N 0 10 V 129 -6 V 0 -12 V 0.15 PolyFill
gsave 5308 2301 N 0 12 V 66 -1 V 0 -15 V 0.15 PolyFill
gsave 5374 2297 N 0 15 V 98 -4 V 0 -16 V 0.15 PolyFill
gsave 5472 2292 N 0 16 V 98 -6 V 0 -16 V 0.15 PolyFill
gsave 5570 2286 N 0 16 V 97 -5 V 0 -16 V 0.15 PolyFill
gsave 5667 2281 N 0 16 V 98 -6 V 0 -15 V 0.15 PolyFill
gsave 5765 2276 N 0 15 V 94 -5 V 0 -15 V 0.15 PolyFill
gsave 5859 2271 N 0 15 V 4 -5 V 0 -10 V 0.15 PolyFill
gsave 5863 2271 N 0 10 V 97 0 V 0 -15 V 0.15 PolyFill
gsave 5960 2266 N 0 15 V 98 -6 V 0 -14 V 0.15 PolyFill
gsave 6058 2261 N 0 14 V 97 -5 V 0 -14 V 0.15 PolyFill
gsave 6155 2256 N 0 14 V 30 -5 V 0 -10 V 0.15 PolyFill
gsave 6185 2255 N 0 10 V 166 -1 V 0 -17 V 0.15 PolyFill
gsave 6351 2247 N 0 17 V 37 -8 V 0 -11 V 0.15 PolyFill
1.000 UL
LTb
1.00 0.00 0.00 C 1175 2490 M
49 9 V
244 29 V
244 10 V
245 -3 V
244 -11 V
244 -15 V
244 -17 V
209 -17 V
35 -2 V
244 -20 V
49 -4 V
78 -6 V
117 -10 V
98 -7 V
98 -8 V
97 -8 V
98 -7 V
98 -8 V
97 -7 V
98 -7 V
97 -7 V
79 -6 V
19 -2 V
98 -6 V
97 -7 V
98 -7 V
98 -6 V
97 -7 V
98 -6 V
98 -6 V
97 -6 V
98 -6 V
129 -8 V
66 -4 V
98 -5 V
98 -6 V
97 -5 V
98 -5 V
94 -5 V
4 0 V
97 -5 V
98 -5 V
97 -5 V
30 -1 V
166 -8 V
37 -2 V
stroke
LTb
1.00 0.00 0.00 C 1175 2379 M
49 18 V
244 64 V
244 46 V
245 1 V
244 -7 V
244 -12 V
244 -12 V
209 -2 V
35 -17 V
244 -4 V
49 -5 V
78 -2 V
117 -7 V
98 -7 V
98 -7 V
97 -7 V
98 -7 V
98 -7 V
97 -7 V
98 -7 V
97 -7 V
79 -6 V
19 -1 V
98 -7 V
97 -7 V
98 -6 V
98 -7 V
97 -6 V
98 -6 V
98 -7 V
97 -6 V
98 -6 V
129 -6 V
66 -1 V
98 -4 V
98 -6 V
97 -5 V
98 -6 V
94 -5 V
4 -5 V
97 0 V
98 -6 V
97 -5 V
30 -5 V
166 -1 V
37 -8 V
1.000 UP
stroke
LT3
0.00 0.50 1.00 C LCb setrgbcolor
LT3
0.00 0.50 1.00 C 6056 4636 M
543 0 V
-543 31 R
0 -62 V
543 62 R
0 -62 V
980 849 M
0 1412 V
949 849 M
62 0 V
949 2261 M
62 0 V
213 -556 R
0 948 V
-31 -948 R
62 0 V
-62 948 R
62 0 V
1468 1028 M
0 970 V
-31 -970 R
62 0 V
-62 970 R
62 0 V
213 -825 R
0 930 V
-31 -930 R
62 0 V
-62 930 R
62 0 V
214 -748 R
0 1065 V
1926 1355 M
62 0 V
-62 1065 R
62 0 V
2201 663 M
0 869 V
2170 663 M
62 0 V
-62 869 R
62 0 V
2445 913 M
0 868 V
2414 913 M
62 0 V
-62 868 R
62 0 V
213 -676 R
0 900 V
-31 -900 R
62 0 V
-62 900 R
62 0 V
2933 816 M
0 896 V
2902 816 M
62 0 V
-62 896 R
62 0 V
213 -75 R
0 1084 V
3146 1637 M
62 0 V
-62 1084 R
62 0 V
3324 682 M
0 1164 V
3293 682 M
62 0 V
-62 1164 R
62 0 V
980 1555 TriU
1224 2179 TriU
1468 1513 TriU
1712 1638 TriU
1957 1888 TriU
2201 1097 TriU
2445 1347 TriU
2689 1555 TriU
2933 1264 TriU
3177 2179 TriU
3324 1264 TriU
6327 4636 TriU
1.000 UP
1.000 UL
LT4
0.00 0.75 0.00 C LCb setrgbcolor
LT4
0.00 0.75 0.00 C 6056 4436 M
543 0 V
-543 31 R
0 -62 V
543 62 R
0 -62 V
3226 1451 M
0 1456 V
3195 1451 M
62 0 V
-62 1456 R
62 0 V
67 -262 R
0 1480 V
3293 2645 M
62 0 V
-62 1480 R
62 0 V
66 -1368 R
0 1505 V
3390 2757 M
62 0 V
-62 1505 R
62 0 V
67 -1322 R
0 1555 V
3488 2940 M
62 0 V
-62 1555 R
62 0 V
67 -1555 R
0 1555 V
3586 2940 M
62 0 V
-62 1555 R
62 0 V
66 -3019 R
0 1572 V
3683 1476 M
62 0 V
-62 1572 R
62 0 V
67 -1293 R
0 1597 V
3781 1755 M
62 0 V
-62 1597 R
62 0 V
67 -1568 R
0 1622 V
3879 1784 M
62 0 V
-62 1622 R
62 0 V
66 -1219 R
0 1314 V
3976 2187 M
62 0 V
-62 1314 R
62 0 V
67 -1701 R
0 1340 V
4074 1800 M
62 0 V
-62 1340 R
62 0 V
4202 794 M
0 1356 V
4171 794 M
62 0 V
-62 1356 R
62 0 V
67 -408 R
0 1373 V
4269 1742 M
62 0 V
-62 1373 R
62 0 V
67 -1760 R
0 1398 V
4367 1355 M
62 0 V
-62 1398 R
62 0 V
66 -907 R
0 1414 V
4464 1846 M
62 0 V
-62 1414 R
62 0 V
67 -1950 R
0 657 V
-31 -657 R
62 0 V
-62 657 R
62 0 V
67 -287 R
0 665 V
-31 -665 R
62 0 V
-62 665 R
62 0 V
66 79 R
0 674 V
-31 -674 R
62 0 V
stroke 4819 2424 M
-62 674 R
62 0 V
67 -957 R
0 907 V
-31 -907 R
62 0 V
-62 907 R
62 0 V
67 -1077 R
0 915 V
-31 -915 R
62 0 V
-62 915 R
62 0 V
66 -1173 R
0 932 V
-31 -932 R
62 0 V
-62 932 R
62 0 V
67 -811 R
0 940 V
-31 -940 R
62 0 V
-62 940 R
62 0 V
67 -1589 R
0 657 V
-31 -657 R
62 0 V
-62 657 R
62 0 V
66 254 R
0 665 V
-31 -665 R
62 0 V
-62 665 R
62 0 V
67 -1048 R
0 765 V
-31 -765 R
62 0 V
-62 765 R
62 0 V
67 -894 R
0 774 V
-31 -774 R
62 0 V
-62 774 R
62 0 V
66 -529 R
0 782 V
-31 -782 R
62 0 V
-62 782 R
62 0 V
67 -1160 R
0 790 V
-31 -790 R
62 0 V
-62 790 R
62 0 V
67 -145 R
0 1081 V
5832 2096 M
62 0 V
-62 1081 R
62 0 V
66 -1585 R
0 1090 V
5929 1592 M
62 0 V
-62 1090 R
62 0 V
67 -1389 R
0 1106 V
6027 1293 M
62 0 V
-62 1106 R
62 0 V
6155 811 M
0 1988 V
6124 811 M
62 0 V
-62 1988 R
62 0 V
3226 2179 CircleF
3324 3385 CircleF
3421 3510 CircleF
3519 3718 CircleF
3617 3718 CircleF
3714 2262 CircleF
3812 2553 CircleF
3910 2595 CircleF
4007 2844 CircleF
4105 2470 CircleF
4202 1472 CircleF
4300 2428 CircleF
4398 2054 CircleF
4495 2553 CircleF
4593 1638 CircleF
4691 2012 CircleF
4788 2761 CircleF
4886 2595 CircleF
4984 2428 CircleF
5081 2179 CircleF
5179 2304 CircleF
5277 1513 CircleF
5374 2428 CircleF
5472 2096 CircleF
5570 1971 CircleF
5667 2220 CircleF
5765 1846 CircleF
5863 2636 CircleF
5960 2137 CircleF
6058 1846 CircleF
6155 1805 CircleF
6327 4436 CircleF
1.000 UP
1.000 UL
LT5
0.55 0.00 0.00 C LCb setrgbcolor
LT5
0.55 0.00 0.00 C 6056 4236 M
543 0 V
-543 31 R
0 -62 V
543 62 R
0 -62 V
6388 1921 M
0 691 V
-31 -691 R
62 0 V
-62 691 R
62 0 V
6185 1517 M
0 667 V
-31 -667 R
62 0 V
-62 667 R
62 0 V
5859 1554 M
0 642 V
-31 -642 R
62 0 V
-62 642 R
62 0 V
5308 1614 M
0 631 V
-31 -631 R
62 0 V
-62 631 R
62 0 V
4281 1923 M
0 562 V
-31 -562 R
62 0 V
-62 562 R
62 0 V
3304 1829 M
0 534 V
-31 -534 R
62 0 V
-62 534 R
62 0 V
2898 2005 M
0 572 V
-31 -572 R
62 0 V
-62 572 R
62 0 V
6388 2266 TriD
6185 1850 TriD
5859 1875 TriD
5308 1929 TriD
4281 2204 TriD
3304 2096 TriD
2898 2291 TriD
6327 4236 TriD
1.000 UL
LT6
1.00 0.00 0.00 C LCb setrgbcolor
LT6
1.00 0.00 0.00 C 6056 4036 M
543 0 V
1175 2440 M
49 13 V
244 45 V
244 20 V
245 5 V
244 -5 V
244 -12 V
244 -15 V
209 -14 V
35 -3 V
244 -18 V
49 -4 V
78 -6 V
20 -1 V
97 -8 V
98 -7 V
98 -8 V
97 -7 V
98 -8 V
98 -7 V
97 -7 V
98 -7 V
97 -7 V
79 -6 V
19 -1 V
98 -7 V
97 -7 V
98 -6 V
98 -7 V
97 -6 V
98 -7 V
98 -6 V
97 -6 V
98 -6 V
98 -6 V
31 -1 V
66 -4 V
98 -6 V
98 -5 V
97 -6 V
98 -5 V
94 -5 V
4 0 V
97 -5 V
98 -5 V
97 -5 V
30 -2 V
166 -8 V
37 -2 V
stroke
LTb
980 4799 N
980 640 L
5859 0 V
0 4159 V
-5859 0 V
Z stroke
1.000 UP
1.000 UL
LTb
stroke
grestore
end
showpage
  }}%
  \put(5936,4036){\makebox(0,0)[r]{\strut{}Reconstruction}}%
  \put(5936,4236){\makebox(0,0)[r]{\strut{}CLEO 2007}}%
  \put(5936,4436){\makebox(0,0)[r]{\strut{}MD-1 1996}}%
  \put(5936,4636){\makebox(0,0)[r]{\strut{}ChrystalBall 1990}}%
  \put(1175,4175){\makebox(0,0)[l]{\strut{}$\bar m_c (pole)=1.85 \pm 0.08 $ GeV}}%
  \put(1175,4591){\makebox(0,0)[l]{\strut{}$\chi^2_{\min}/\text{d.o.f}=1.03$}}%
  \put(3909,140){\makebox(0,0){\strut{}$\sqrt{s}$ (in GeV)}}%
  \put(160,2719){\makebox(0,0){\strut{}$R_{exp.}$, $\textcolor{red}{R_{th.}}$}}%
  \put(6839,440){\makebox(0,0){\strut{} 11}}%
  \put(5863,440){\makebox(0,0){\strut{} 10}}%
  \put(4886,440){\makebox(0,0){\strut{} 9}}%
  \put(3910,440){\makebox(0,0){\strut{} 8}}%
  \put(2933,440){\makebox(0,0){\strut{} 7}}%
  \put(1957,440){\makebox(0,0){\strut{} 6}}%
  \put(980,440){\makebox(0,0){\strut{} 5}}%
  \put(860,4799){\makebox(0,0)[r]{\strut{} 4.2}}%
  \put(860,4383){\makebox(0,0)[r]{\strut{} 4.1}}%
  \put(860,3967){\makebox(0,0)[r]{\strut{} 4}}%
  \put(860,3551){\makebox(0,0)[r]{\strut{} 3.9}}%
  \put(860,3135){\makebox(0,0)[r]{\strut{} 3.8}}%
  \put(860,2720){\makebox(0,0)[r]{\strut{} 3.7}}%
  \put(860,2304){\makebox(0,0)[r]{\strut{} 3.6}}%
  \put(860,1888){\makebox(0,0)[r]{\strut{} 3.5}}%
  \put(860,1472){\makebox(0,0)[r]{\strut{} 3.4}}%
  \put(860,1056){\makebox(0,0)[r]{\strut{} 3.3}}%
  \put(860,640){\makebox(0,0)[r]{\strut{} 3.2}}%
\end{picture}%
\endgroup
 